\documentclass[amsmath,amssymb,aps,pra,10pt,twocolumn,groupedaddress,nofootinbib]{revtex4-1}

\usepackage{graphicx, epsf} 
\usepackage{color} 
\usepackage[table]{xcolor}
\usepackage{dcolumn} 
\usepackage{import} 
\usepackage[skins]{tcolorbox} 
\usepackage{titlesec, setspace} 
\usepackage{listings} 
\usepackage[T1]{fontenc} 
\usepackage{caption} 
\usepackage[bitstream-charter]{mathdesign} 
\graphicspath{{figures/}}
\input{Qcircuit}

\usepackage{calrsfs}
\DeclareMathAlphabet{\pazocal}{OMS}{zplm}{m}{n}
\let\mathcal\undefined
\newcommand{\mathcal}[1]{\pazocal{#1}}

\let\bm\undefined
\newcommand{\bm}[1]{\mathbf{#1}}

\usepackage[newfloat, frozencache]{minted}

\usepackage{tabularx,booktabs}
\usepackage{cellspace}
\usepackage{placeins}

\newcolumntype{L}[1]{>{\raggedright\arraybackslash}p{#1}}
\arrayrulecolor{white}
\usepackage[hyperfootnotes=true,colorlinks=true,linkcolor=magenta]{hyperref}

\renewcommand{\Qcircuit}[1][0em]{\xymatrix @*=<#1>}

\newcommand{\I}{\mathbb{I}}
\renewcommand{\Re}{\mathrm{Re}}
\renewcommand{\Im}{\mathrm{Im}}
\newcommand{\tr}{\mathrm{Tr}}
\renewcommand{\a}{\hat{a}}

\newcommand{\ad}{\a^\dagger}
\newcommand{\n}{\hat{n}}

\newcommand{\x}{\hat{x}}
\newcommand{\p}{\hat{p}}
\newcommand{\nbar}{\overline{n}}

\newcommand{\ketbra}[2]{| #1 \rangle\langle #2 |}

\newcommand{\Rgate}{\Qcircuit @C=0.5em @R=.7em {& \gate{R} & \qw}}
\newcommand{\Dgate}{\Qcircuit @C=0.5em @R=.7em {& \gate{D} & \qw}}
\newcommand{\Sgate}{\Qcircuit @C=0.5em @R=.7em {& \gate{S} & \qw}}
\newcommand{\BSgate}{\Qcircuit @C=0.5em @R=.7em {
& \multigate{1}{BS} & \qw \\
& \ghost{BS}& \qw 
}}
\newcommand{\Vgate}{\Qcircuit @C=0.5em @R=.7em {& \gate{V} & \qw}}

\definecolor{codegreen}{rgb}{0,0.6,0}
\definecolor{codegray}{rgb}{0.5,0.5,0.5}
\definecolor{codepurple}{rgb}{0.58,0,0.82}
\definecolor{backcolour}{RGB}{230, 240, 230}
\definecolor{BlueChill}{RGB}{32, 147, 149}
\definecolor{Shamrock}{RGB}{50, 211, 167}
\titleformat{\section}[display]{\vspace{-1em}}{}{0pt}{\normalfont\sffamily\color{BlueChill}}[\vspace{-3pt}\hrule]
\titleformat{\subsection}[display]{\vspace{-2.5em}}{}{0pt}{\normalfont\sffamily\color{BlueChill}}[\vspace{-1em}]
\titleformat{\subsubsection}[display]{\vspace{-2.5em}}{}{0pt}{\bfseries\color{black}}[\vspace{-1em}]
\makeatletter
\def\frontmatter@abstractfont{\sffamily\color{black}\setstretch{1.5}}%
\def\frontmatter@title@format{\noindent\huge\sffamily\color{BlueChill}}{}%
\def\frontmatter@authorformat{\vspace{1em}\noindent\color{BlueChill}\Large\sffamily}%
\def\frontmatter@affiliationfont{\vspace{1em}\color{BlueChill}\noindent\normalsize\sffamily}%
\def\frontmatter@above@affiliation@script{\vspace{1em}\noindent}%
\def\frontmatter@makefnmark{}
\renewcommand*\frontmatter@date[2][\Dated@name]{\def\@date{}}%
\makeatother

\lstdefinestyle{codeblock}{
  backgroundcolor=\color{backcolour},   
  commentstyle=\color{codegreen},
  keywordstyle=\color{blue},
  numberstyle=\tiny\color{codegray},
  stringstyle=\color{codepurple},
  basicstyle=\footnotesize,
  escapechar=\¢,escapebegin=\color{purple}, 
  otherkeywords={with},
  breakatwhitespace=false,         
  breaklines=true,                 
  captionpos=b,                    
  keepspaces=true,
  language=Python,
  numbers=right,                    
  numbersep=5pt,                  
  showspaces=false,                
  showstringspaces=false,
  showtabs=false,                  
  tabsize=2,
  basicstyle=\ttfamily\footnotesize
}
\definecolor{lightgreen}{HTML}{EDF7F4}
\definecolor{xgreen}{HTML}{119A68}
\titleformat{\section}[display]{\vspace{-1em}}{}{0pt}{\Large}[\vspace{-3pt}\color{xgreen}\hrule]
\titleformat{\subsection}[display]{\vspace{-2.5em}}{}{0pt}{\bfseries\color{xgreen}}[\vspace{-1em}]
\titleformat{\subsubsection}[display]{\vspace{-2.5em}}{}{0pt}{\itshape\color{black}}[\vspace{-1em}]

\makeatletter
\def\frontmatter@abstractfont{\color{black}\setstretch{1.5}}%
\def\frontmatter@title@format{\noindent\huge\color{black}}{}%
\def\frontmatter@authorformat{\vspace{1em}\noindent\color{xgreen}\Large}%
\def\frontmatter@affiliationfont{\vspace{1em}\itshape\color{black}\noindent\normalsize}%
\def\frontmatter@above@affiliation@script{\vspace{1em}\noindent}%
\def\frontmatter@makefnmark{}
\renewcommand*\frontmatter@date[2][\Dated@name]{\def\@date{}}%
\makeatother

\usepackage[format=plain,labelfont={bf,it},textfont=it]{caption}
\newenvironment{code}{\captionsetup{type=listing}}{}
\SetupFloatingEnvironment{listing}{name=Codeblock}
\newmintedfile[pythonfile]{python}{
	bgcolor=gray!6,
	fontfamily=tt,
	gobble=0,
	fontsize=\small,
	breaklines=true,
	framerule=0.4pt,
	funcnamehighlighting=true,
	tabsize=2,
	obeytabs=false,
	mathescape=false
	samepage=false, 
	showspaces=false,
	showtabs =false,
	texcl=false,
	linenos=true,
	xleftmargin=1pt,
	numbersep=1pt
}

\usepackage{float}
\floatstyle{ruled}
\newfloat{floatbox}{htp}{box}
\floatname{floatbox}{Box}
\newcommand{\infobox}[2]{
    \begin{floatbox}
        \caption{#1}
        \let\centering\relax
        \parbox{.94\columnwidth}{#2}
    \end{floatbox}
}

\usepackage{silence}
\begin{document}
\title{Strawberry Fields:\\ A Software Platform for Photonic Quantum Computing}

\begin{abstract}
We introduce Strawberry Fields, an open-source quantum programming architecture for light-based quantum computers, and detail its key features. Built in Python, Strawberry Fields is a full-stack library for design, simulation, optimization, and quantum machine learning of continuous-variable circuits. 
The platform consists of three main components: (i) an API for quantum programming based on an easy-to-use language named Blackbird; (ii) a suite of three virtual quantum computer backends, built in NumPy and TensorFlow, each targeting specialized uses; and (iii) an engine which can compile Blackbird programs on various backends, including the three built-in simulators, and -- in the near future -- photonic quantum information processors.
The library also contains examples of several paradigmatic algorithms, including teleportation, (Gaussian) boson sampling, instantaneous quantum polynomial, Hamiltonian simulation, and variational quantum circuit optimization.
\end{abstract}

\author{Nathan Killoran, Josh Izaac, Nicol\'{a}s Quesada, Ville Bergholm, Matthew Amy, and Christian Weedbrook}
\affiliation{Xanadu, 372 Richmond St W, Toronto, M5V 1X6, Canada}
\maketitle

\section{Introduction}
The decades-long worldwide quest to build practical quantum computers is currently undergoing a critical period. During the next few years, a number of different quantum devices will become available to the public. While fault-tolerant quantum computers will one day provide significant computational speedups for problems like factoring \cite{shor1999polynomial}, search \cite{grover1996fast}, or linear algebra \cite{harrow2009quantum}, near-term quantum devices will be noisy, approximate, and sub-universal \cite{preskill2018quantum}. Nevertheless, these emerging quantum processors are expected to be strong enough to show a computational advantage over classical computers for certain problems, an achievement known as quantum computational supremacy. 

As we approach this milestone, work is already underway exploring how such quantum processors might best be leveraged. Popular techniques include variational quantum eigensolvers \cite{peruzzo2014variational, mcclean2016theory}, quantum approximate optimization algorithms \cite{farhi2014quantum, farhi2016quantum}, sampling from computationally hard probability distributions 
\cite{aaronson2011computational, hamilton2017gaussian, chakhmakhchyan2017boson, bremner2010classical, bremner2016average, boixo2016characterizing, aaronson2016complexity, neill2017blueprint, douce2017iqp}, 
and quantum annealing \cite{finnila1994quantum, johnson2011quantum}. Notably, these methodologies can be applied to tackle important practical problems in chemistry \cite{yung2014transistor, omalley2016scalable,  shen2017quantum, kandala2017hardware}, finance \cite{rosenberg2016solving}, optimization \cite{lucas2014ising, neukart2017traffic}, and machine learning \cite{neven2009training, pudenz2013quantum, crawford2016reinforcement, amin2016quantum, riste2017demonstration, verdon2017quantum, otterbach2017unsupervised, schuld2018quantum}. 
These known applications are very promising, yet it is perhaps the \emph{unknown} future applications of quantum computers that are most tantalizing. We may not know the best applications of quantum computers until these devices become available more widely to researchers, students, entrepreneurs, and programmers worldwide. 

To this end, a nascent quantum software ecosystem has recently begun to develop \cite{green2013quipper, wecker2014liqui, javadiabhari2015scaffcc, smelyanskiy2016qhipster, pakin2016quantum, smith2016practical, steiger2016projectq, cross2017open, fried2017qtorch, mccaskey2017extreme, mcclean2017openfermion, liu2017q, svore2018q}. 
However, a prevailing theme for these software efforts is to target qubit-based quantum devices. In reality, there are several competing models of quantum computing which are equivalent in computational terms, but which are conceptually quite distinct. One prominent approach is the continuous variable (CV) model of quantum computing \cite{lloyd1999quantum, braunstein2005quantum, weedbrook2012gaussian}. In the CV model, the basic information-processing unit is an infinite-dimensional bosonic mode, making it particularly well-suited for implementations and applications based on light. The CV model retains the computational power of the qubit model (cf. Chap. 4 of Ref. \cite{nielsen2002quantum}), while offering a number of unique features. For instance, the CV model is a natural fit for simulating bosonic systems (electromagnetic fields, trapped atoms, harmonic oscillators, Bose-Einstein condensates, phonons, or optomechanical resonators) and for settings where continuous quantum operators -- such as position \& momentum -- are present. Even in classical computing, recent advances from deep learning have demonstrated the power and flexibility of a continuous representation of computation \cite{graves2014neural, graves2016hybrid} in comparison to the discrete computational model which has historically dominated.

Here we introduce Strawberry Fields\footnote{This document refers to Strawberry Fields version 0.9. Full documentation is available online at \href{https://strawberryfields.readthedocs.io}{strawberryfields.readthedocs.io}.}, an open-source software architecture for photonic quantum computing. Strawberry Fields is a full-stack quantum software platform, implemented in Python, specifically targeted to the CV model. 
Its main element is a new quantum programming language named Blackbird. To lay the groundwork for future photonic quantum computing hardware, Strawberry Fields also includes a suite of three CV quantum simulator backends implemented using NumPy \cite{walt2011numpy} and TensorFlow \cite{abadi2016tensorflow}. 
Strawberry Fields comes with a built-in engine to convert Blackbird programs to run on any of the simulator backends or, when they are available, on photonic quantum computers. To accompany the library, an online service for interactive exploration and simulation of CV circuits is available at \href{https://strawberryfields.ai}{strawberryfields.ai}.

Aside from being the first quantum software framework to support photonic quantum computation with continuous-variables, Strawberry Fields provides additional computational features not presently available in the quantum software ecosystem:
\begin{enumerate}
	\item  We provide two numeric simulators; a Gaussian backend, and a Fock-basis backend. These two formulations are unique to the CV model of quantum computation due to the use of an infinite Hilbert space, and came with their own technical challenges.
	\begin{enumerate}
		\item The Gaussian backend provides state-of-the-art methods and functions for calculating the fidelity and Fock state probabilities, involving calculations of the classically intractable hafnian \cite{quesada2018faster}.
		\item The Fock backend allows operations such as squeezing and beamsplitters to be performed in the Fock-basis, a computationally intensive calculation that has been highly vectorized and benchmarked for performance.
	\end{enumerate}
	\item We provide a suite of important circuit decompositions appearing in quantum photonics -- such as the Williamson, Bloch-Messiah, and Clements decompositions. 
	\item The Fock-basis backend written using the TensorFlow machine learning library allows for symbolic calculations, automatic differentiation, and backpropagation through CV quantum simulations. As far as we are aware, this is the first quantum simulation library written using a high-level machine learning library, with support for dataflow programming and automatic differentiation.
\end{enumerate}

The remainder of this white paper is structured as follows. Before presenting Strawberry Fields, we first provide a brief overview of the key ingredients for CV quantum computation, specifically the most important states, gates, and measurements. We then introduce the Strawberry Fields architecture in full, presenting the Blackbird quantum assembly language, outlining how to use the library for numerical simulation, optimization, and quantum machine learning. Finally, we discuss the three built-in simulators and the internal representations that they employ. In the Appendices, we give further mathematical and software details and provide full example code for a number of important CV quantum computing tasks. 

\section{Quantum Computation with Continuous Variables}

Many physical systems in nature are intrinsically continuous, with light being the prototypical example. Such systems reside in an infinite-dimensional Hilbert space, offering a paradigm for quantum computation which is distinct from the discrete qubit model. This \emph{continuous-variable model} takes its name from the fact that the quantum operators underlying the model have continuous spectra. It is possible to embed qubit-based computations into the CV picture \cite{gottesman2001encoding}, so the CV model is as powerful as its qubit counterparts. 

\subsection{From Qubits to Qumodes}

A high-level comparison of CV quantum computation with the qubit model is depicted in Table \ref{box:cv_vs_qubits}. In the remainder of this section, we will provide a basic presentation of the key elements of the CV model. A more detailed technical overview can be found in Appendix \ref{app:technical_details}. Readers experienced with CV quantum computing can safely skip to the next section.

\begin{table}[h]
    \centering
    \small
    \rowcolors{1}{gray!10}{white}
    \arrayrulecolor{xgreen} 
    \setlength{\tabcolsep}{0pt}
    \setlength\extrarowheight{5pt}
      \begin{tabular}{m{2.3cm} >{\raggedright\arraybackslash}m{3.2cm}  >{\raggedright\arraybackslash}m{3cm}}
      	\rowcolor{lightgreen}
	                 & \textbf{CV}                                 & \textbf{Qubit} \\ 
	\toprule
	Basic element      & Qumodes                     & Qubits \\
	Relevant operators & Quadratures $\x,\p$\newline Mode operators $\a,\ad$  &  Pauli operators $\hat{\sigma}_x, \hat{\sigma}_y, \hat{\sigma}_z$\\
	Common states      & Coherent states $\ket{\alpha}$ Squeezed states $\ket{z}$  Number states $\ket{n}$    & Pauli eigenstates $\ket{0/1}, \ket{\pm}, \ket{\pm i}$\\
	Common gates       & Rotation,\newline Displacement, Squeezing, Beamsplitter, Cubic Phase & Phase shift, Hadamard, CNOT, T-Gate \\
	Common measurements & Homodyne $\ketbra{x_\phi}{x_\phi}$, Heterodyne $\tfrac{1}{\pi}\ketbra{\alpha}{\alpha}$,  Photon-counting $\ketbra{n}{n}$    & Pauli eigenstates $\ketbra{0/1}{0/1}, \ketbra{\pm}{\pm}$, $\ketbra{\pm i}{\pm i}$
      \end{tabular}
  \captionsetup{name=Table,type=table}
  \caption{Basic comparison of the CV and qubit settings.}
  \label{box:cv_vs_qubits}
\end{table}

The most elementary CV system is the bosonic harmonic oscillator, defined via the canonical mode operators $\a$ and $\ad$. These satisfy the well-known commutation relation $[\a,\ad]=\I$. It is also common to work with the \emph{quadrature operators} (also called the position \& momentum operators)\footnote{It is common to picture $\hbar$ as a (dimensionless) scaling parameter for the $\x$ and $\p$ operators rather than a physical constant. However, there are several conventions for the scaling value in common use \cite{ferraro2005gaussian}. These self-adjoint operators are proportional to the Hermitian and anti-Hermitian parts of the operator $\a$. 
Strawberry Fields allows the user to specify this value, with the default $\hbar=2$.},
\begin{align}
 \hat{x} := \sqrt{\frac{\hbar}{2}}(\a + \ad), \\
 \hat{p} :=  -i \sqrt{\frac{\hbar}{2}}(\a - \ad),
\end{align}
where $[\x,\p]=i \hbar \hat{\I}$.
We can picture a fixed harmonic oscillator mode (say, within an optical fibre or a waveguide on a photonic chip) as a single `wire' in a quantum circuit. These \emph{qumodes} are the fundamental information-carrying units of CV quantum computers. By combining multiple qumodes -- each with corresponding operators $\a_i$ and $\ad_i$ -- and interacting them via sequences of suitable quantum gates, we can implement a general CV quantum computation.


\subsection{CV States}

The dichotomy between qubit and CV systems is perhaps most evident in the basis expansions of quantum states:
\begin{align}
  &\rm{Qubit} &\ket{\phi} & = \phi_0 \ket{0} + \phi_1 \ket{1}, \\
  &\rm{Qumode} &\ket{\psi} & = \int dx~\psi(x) \ket{x}. \label{eq:dichotomy}
\end{align}
For qubits, we use a discrete set of coefficients; for CV systems, we can have a \emph{continuum}. The states $\ket{x}$ are the eigenstates of the $\x$ quadrature, $\x\ket{x}=x\ket{x}$, with $x\in\mathbb{R}$. These quadrature states are special cases of a more general family of CV states, the \emph{Gaussian states}, which we now introduce. 

\subsubsection{Gaussian states}
\begin{figure}[t]
\begin{center}
\includegraphics[width=0.6\columnwidth]{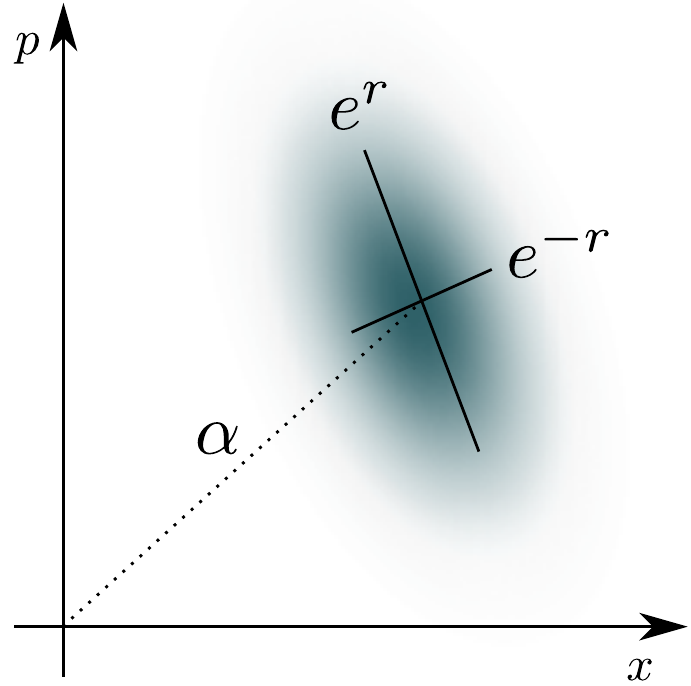}
\end{center}
\caption{Schematic representation of a Gaussian state for a single mode. The shape and orientation are parameterized by the displacement $\alpha$ and squeezing $z=r\exp(i\phi)$.}
\label{fig:gaussian}
\end{figure}

Our starting point is the vacuum state $\ket{0}$. Other states can be created by evolving the vacuum state according to 
\begin{equation}
 \label{eq:gaussian_state_creation}
 \ket{\psi} = \exp(-itH)\ket{0},
\end{equation}
where $H$ is a bosonic Hamiltonian (i.e., a function of the operators $\a_i$ and $\ad_i$) and $t$ is the evolution time. States where the Hamiltonian $H$ is at most quadratic in the operators $\a_i$ and $\ad_i$ (equivalently, in $\x_i$ and $\p_i$) are called \emph{Gaussian}. 
For a single qumode, Gaussian states are parameterized by two continuous complex variables: a displacement parameter $\alpha\in\mathbb{C}$ and a squeezing parameter $z\in\mathbb{C}$ (often expressed as $z=r\exp(i\phi)$, with $r \geq 0$). Gaussian states are so-named because we can identify each Gaussian state with a corresponding Gaussian distribution. For single qumodes, the identification proceeds through its displacement and squeezing parameters. 
The displacement gives the centre of the distribution, while the squeezing determines the variance and rotation of the distribution (see Fig. \ref{fig:gaussian}).
Multimode Gaussian states, on the other hand, are parameterized by a vector of displacements $\bar{ \bm{r}}$ and a covariance matrix $\mathbf{V}$.
Many important pure states in the CV model are special cases of the pure Gaussian states; see Table \ref{tab:states} for a summary. 

\begin{table}[h]
    \centering
    \small
	\rowcolors{1}{gray!10}{white}
	\arrayrulecolor{xgreen} 
	\setlength{\tabcolsep}{0pt}
	\setlength\extrarowheight{5pt}
	\begin{tabular}{
		m{3.2cm} 
		>{\raggedright\arraybackslash}m{2.5cm} 
		>{\raggedright\arraybackslash}m{2.7cm}
	      }
	    \rowcolor{lightgreen}
		\textbf{State family     }              & \textbf{Displacement}                      & \textbf{Squeezing} \\
		\toprule
		Vacuum state $\ket{0}$         & $\alpha=0$                        & $z=0$ \\
		Coherent states $\ket{\alpha}$ & $\alpha\in\mathbb{C}$             & $z=0$ \\
		Squeezed states $\ket{z}$      & $\alpha=0$                        & $z\in\mathbb{C}$ \\ 
		Displaced squeezed\newline states $\ket{\alpha,z}$            & $\alpha\in\mathbb{C}$             & $z\in\mathbb{C}$ \\
		$\x$ eigenstates $\ket{x}$     & $\alpha\in\mathbb{C}$, $x=2 \sqrt{\frac{\hbar}{2}}\mathrm{Re}(\alpha)$             & $\phi=0$, $r\rightarrow\infty$ \\
		$\p$ eigenstates $\ket{p}$     & $\alpha\in\mathbb{C}$, $p=2 \sqrt{\frac{\hbar}{2}}\mathrm{Im}(\alpha)$             & $\phi=\pi$, $r\rightarrow\infty$ \\
		Fock states $\ket{n}$          & N.A.                              & N.A.
      \end{tabular}
  \captionsetup{name=Table,type=table}
  \caption{Common single-mode pure states and their relation to the displacement and squeezing parameters. All listed families are Gaussian, except for the Fock states. The $n=0$ Fock state is also the vacuum state.}
  \label{tab:states}
\end{table}

\subsubsection{Fock states}

Complementary to the continuous Gaussian states are the discrete \emph{Fock states} (or \emph{number states}) $\ket{n}$, where $n$ are nonnegative integers. 
These are the eigenstates of the number operator $\n=\ad\a$.
The Fock states form a discrete (countable) basis for qumode systems. Thus, each of the Gaussian states considered in the previous section can be expanded in the Fock-state basis. For example, coherent states have the form
\begin{equation}
\label{eq:coherentstate}
 \ket{\alpha} = \exp\left(-\tfrac{|\alpha|^2}{2}\right) \sum_{n=0}^\infty \frac{\alpha^n}{\sqrt{n!}}\ket{n},
\end{equation}
while (undisplaced) squeezed states only have even number states in their expansion:
\begin{equation}
\label{eq:squeezedstate}
 \ket{z} = \frac{1}{\sqrt{\cosh r}}\sum_{n=0}^\infty\frac{\sqrt{(2n)!}}{2^n n!}[-e^{i\phi}\tanh (r)]^n\ket{2n}.
\end{equation}

\subsubsection{Mixed states}

Mixed Gaussian states are also important in the CV picture, for instance, the \emph{thermal state}
\begin{equation}
\label{eq:thermalstate}
 \rho(\nbar) := \sum_{n=0}^\infty\frac{\nbar^n}{(1+\nbar)^{n+1}}\ketbra{n}{n},
\end{equation}
which is parameterized via the mean photon number $\nbar:=\tr{(\rho(\nbar)\hat{n})}$. Starting from this state, we can consider a mixed-state-creation process similar to Eq. (\ref{eq:gaussian_state_creation}), namely
\begin{equation}
 \label{eq:gaussian_mixed_state_creation}
 \rho = \exp(-itH)\rho(\nbar)\exp(itH).
\end{equation}
Analogously to pure states, by applying Hamiltonians of second-order (or lower) to thermal states, we generate the family of Gaussian mixed states.


\subsection{CV Gates}

Unitary operations can be associated with a generating Hamiltonian $H$ via the recipe (cf. Eqs. (\ref{eq:gaussian_state_creation}) \& (\ref{eq:gaussian_mixed_state_creation}))
\begin{equation}
 \label{eq:unitary_generator}
 U := \exp{(-itH)}.
\end{equation}
For convenience, we classify unitaries by the degree of their generator.
A CV quantum computer is said to be universal if it can implement, to arbitrary precision and with a finite number of steps, any unitary which is polynomial in the mode operators \cite{lloyd1999quantum}. 
We can build a multimode unitary by applying a sequence of gates from a \emph{universal gate set}, each of which acts only on one or two modes. We focus on a universal set made from the following two subsets: 
\begin{description}
 \item [Gaussian gates] Single and two-mode gates which are at most quadratic in the mode operators, e.g., \emph{Displacement, Rotation, Squeezing, and Beamsplitter} gates.
 \item [Non-Gaussian gate] A single-mode gate which is degree 3 or higher, e.g., the \emph{Cubic phase} gate.
\end{description}

A number of fundamental CV gates are presented in Table \ref{box:gates}. 
Many of the Gaussian states from the previous section are connected to a corresponding Gaussian gate. Any multimode Gaussian gate can be implemented through a suitable combination of Displacement, Rotation, Squeezing, and Beamsplitter Gates \cite{weedbrook2012gaussian}, making these gates sufficient for constructing all quadratic unitaries. The cubic phase gate is presented as an exemplary non-Gaussian gate, but any other non-Gaussian gate could also be used to achieve universality. 
A number of other useful CV gates are listed in Appendix \ref{app:gates_etc}.

\begin{table}[ht]
	\small
	\centering
	\rowcolors{1}{gray!10}{white}
	\arrayrulecolor{xgreen} 
	\setlength{\tabcolsep}{0pt}
	\setlength\extrarowheight{5pt}
	\begin{tabular}{
			m{2.2cm}
			>{\raggedright\arraybackslash}m{4.2cm}
			>{\centering\arraybackslash}m{2cm}
		}
		\rowcolor{lightgreen}
		\textbf{Gate}         & \textbf{Unitary}                                                 & \textbf{Symbol} \\ 
		\toprule
		Displacement & $D_i(\alpha)=\exp{(\alpha\ad_i - \alpha^*\a_i)}$              & \parbox{\textwidth}{\Dgate} \\[5pt]
		Rotation     & $R_i(\phi)=\exp{(i\phi\hat{n}_i)}$                          &\parbox{\textwidth}{\Rgate} \\[5pt]
		Squeezing    & $S_i(z)=\exp{(\frac{1}{2}(z^* \a_i^2 - z \a_i^{\dagger 2}))}$ & \parbox{\textwidth}{\Sgate} \\[5pt]
		Beamsplitter & $BS_{ij}(\theta,\phi)=\exp{(\theta(e^{i\phi}\a_i\ad_j - e^{-i\phi}\ad_i\a_j))}$                                      & \parbox{\textwidth}{\BSgate} \\[5pt]
		Cubic phase  & $V_i(\gamma)=\exp{\left(i\frac{\gamma}{3 \hbar }\x_i^3\right)}$               & \parbox{\textwidth}{\Vgate}
      \end{tabular}
  \captionsetup{name=Table,type=table}
  \caption{Some important CV model gates. All listed gates except the cubic phase gate are Gaussian.}
  \label{box:gates}
\end{table}


\subsection{CV Measurements}

\begin{table}[t]
	\small
	\centering
	\rowcolors{1}{gray!10}{white}
	\arrayrulecolor{xgreen} 
	\setlength{\tabcolsep}{0pt}
	\setlength\extrarowheight{5pt}
	\begin{tabular}{
		m{3cm}
		>{\centering\arraybackslash}m{3.2cm} 
		>{\centering\arraybackslash}m{2.38cm}
	}
		\rowcolor{lightgreen}
		\textbf{Measurement}      & \textbf{Measurement Operators  }                   & \textbf{Measurement values} \\
		\toprule
		Homodyne         & $\ketbra{x_\phi}{x_\phi}$                 & $x\in\mathbb{R}$ \\[5pt]
		Heterodyne       & $\frac{1}{\pi}\ketbra{\alpha}{\alpha}$ & $\alpha\in\mathbb{C}$ \\[5pt]
		Photon counting  & $\ketbra{n}{n}$                           & $n\in\mathbb{N}$
  \end{tabular}
  \captionsetup{name=Table,type=table}
  \caption{Key measurement types for the CV model. The `-dyne' measurements are Gaussian, while photon-counting is non-Gaussian.}
  \label{box:meas}
\end{table}

As with CV states and gates, we can distinguish between Gaussian and non-Gaussian measurements. The Gaussian class consists of two (continuous) types: homodyne and heterodyne measurements, while the key non-Gaussian measurement is photon counting. These are summarized in Table \ref{box:meas}.

\subsubsection{Homodyne measurements}

Ideal homodyne detection is a projective measurement onto the eigenstates of the quadrature operator $\x$. These states form a continuum, so homodyne measurements are inherently continuous, returning values $x\in\mathbb{R}$.
More generally, we can consider projective measurement onto the eigenstates $\ket{x_\phi}$ of the Hermitian operator
\begin{equation}
 \x_\phi:=\cos\phi~\x + \sin\phi~\p.
\end{equation}
This is equivalent to rotating the state clockwise by $\phi$ and performing an $\x$-homodyne measurement. 
If we have a multimode Gaussian state and we perform homodyne measurement on one of the modes, the conditional state of the unmeasured modes remains Gaussian. 

\subsubsection{Heterodyne measurements}

Whereas homodyne detection is a measurement of $\x$, heterodyne detection can be seen as a simultaneous measurement of both $\x$ and $\p$. Because these operators do not commute, they cannot be simultaneously measured without some degree of uncertainty. Equivalently, we can picture heterodyne measurement as projection onto the coherent states, with measurement operators $\frac{1}{\pi}\ketbra{\alpha}{\alpha}$. Because the coherent states are not orthogonal, there is a corresponding lack of sharpness in the measurements.
If we perform heterodyne measurement on one mode of a multimode state, the conditional state on the remaining modes stays Gaussian. 

\subsubsection{Photon Counting}
Photon counting (also known as as \emph{photon-number resolving measurement}), is a complementary measurement method to the `-dyne' measurements, revealing the particle-like, rather than the wave-like, nature of qumodes. 
This measurement projects onto the number eigenstates $\ket{n}$, returning non-negative integer values $n\in\mathbb{N}$. Except for the outcome $n=0$, a photon-counting measurement on a single mode of a multimode Gaussian state will cause the remaining modes to become non-Gaussian. Thus, photon-counting can be used as an ingredient for implementing non-Gaussian gates. A related process is \emph{photodetection}, where a detector only resolves the vacuum state from non-vacuum states. This process has only two measurement operators, namely $\ketbra{0}{0}$ and $\I - \ketbra{0}{0}$.



\section{The Strawberry Fields Software Platform}

\begin{figure*}[th]
\begin{center}
\includegraphics[width=0.95\linewidth]{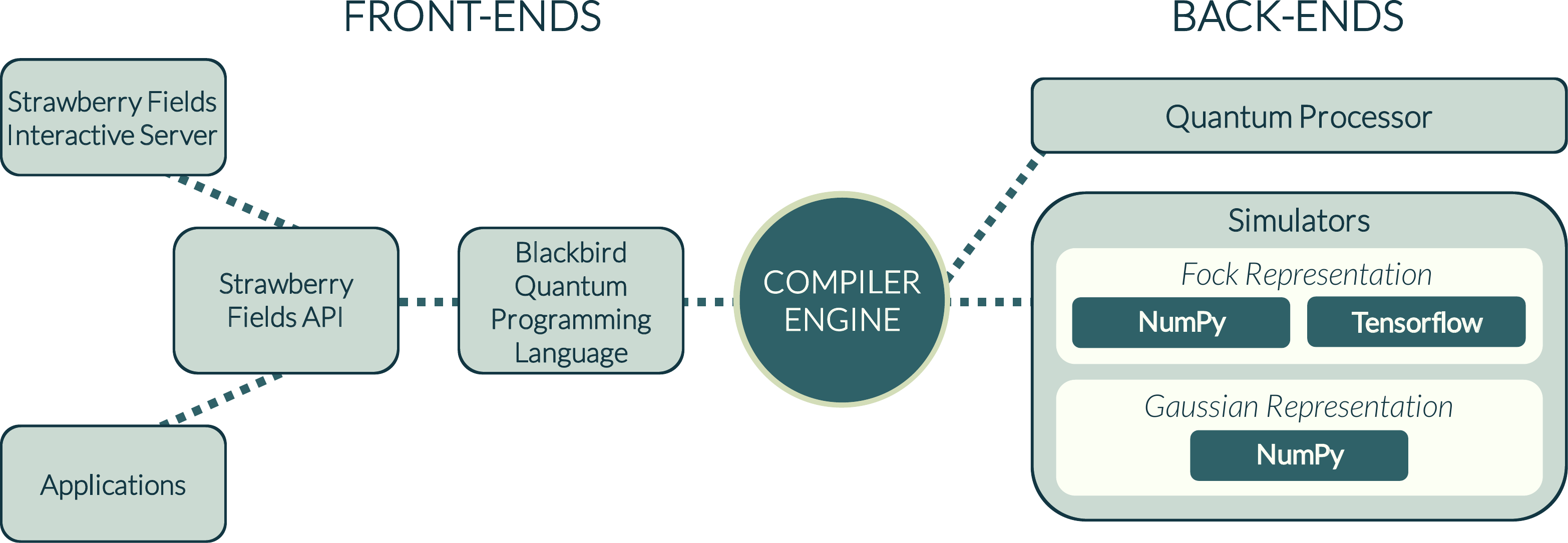}
\end{center}
\caption{Outline of the Strawberry Fields software stack. The Strawberry Fields Interactive server is available online at \href{https://strawberryfields.ai}{strawberryfields.ai}.}
\label{fig:sf_stack}
\end{figure*}

The Strawberry Fields library has been designed with several key goals in mind. Foremost, it is a standard-bearer for the CV model, laying the groundwork for future photonic quantum computers. As well, Strawberry Fields is designed to be simple to use, giving entry points for as many users as possible. 
Finally, since the potential applications of near-term quantum computers are still being worked out, it is important that Strawberry Fields provides powerful tools to easily explore many different use-cases and applications. 

Strawberry Fields has been implemented in Python, a modern language with a gentle learning curve which is already familiar to many programmers and scientific practitioners. The accompanying quantum simulator backends are built upon the widely used Python packages NumPy and TensorFlow. All Strawberry Fields code is open source. Strawberry Fields can be accessed programmatically as a Python package, or via a browser-based interface for designing quantum circuits.

A pictorial outline of Strawberry Fields' key elements and their interdependencies is presented in Fig. \ref{fig:sf_stack}. Conceptually, the software stack is separated into two main pieces: a user-facing frontend layer and a lower-level backends component. The frontend encompasses the Strawberry Fields Python API and the Blackbird quantum assembly language. These elements provide access points for users to design quantum circuits. These circuits are then linked to a backend via a quantum compiler engine. For a backend, the engine can currently target one of three included quantum computer simulators. When CV quantum processors become available in the near future, the engine will also build and run circuits on those devices. Further, high-level quantum computing applications can be built by leverging the Strawberry Fields frontend API. Existing examples include the Strawberry Fields Interactive website, the Quantum Machine Learning Toolbox (for streamlining the training of variational quantum circuits), and SFOpenBoson (an interface between the electronic structure library OpenFermion \cite{mcclean2017openfermion} and Strawberry Fields).

In the remainder of this section, the key elements of Strawberry Fields will be presented in more detail. Proceeding through a series of examples, we show how CV quantum computations can be defined using the Blackbird language, then compiled and run on a quantum computer backend. We also outline how to use Strawberry Fields for optimization and machine learning on quantum circuits. Finally, we discuss the suite of quantum computer simulators included within Strawberry Fields.


\subsection{Blackbird: A Quantum Programming Language}

As classical computers have become progressively more powerful, the languages used to program them have also undergone considerable paradigmatic changes. Machine code gave way to human-readable assembly languages, followed by higher-level procedural and object-oriented languages. With each generation, the trend has been towards higher levels of abstraction, separating the programmer more and more from details of the actual computer hardware. Quantum computers are still at an early stage of development, so while we can imagine what higher-level quantum programming might look like, in the near term we first need to build languages which are conceptually closer to the quantum hardware. 

Blackbird is a standalone domain specific language (DSL) for continuous-variable quantum computation. With a well-defined grammar in extended Backus-Naur form, and both Python and C++ parsers available, Blackbird provides operations that match the basic CV states, gates, and measurements, and maps directly to low-level hardware instructions. The abstract syntax keeps a close connection between code and the quantum operations that they implement; this syntax is modeled after that of ProjectQ \cite{steiger2016projectq}, but specialized to the CV setting. Blackbird can be used as part of the Strawberry Fields stack, but also directly with photonic quantum computing hardware systems. 
	
Within the Strawberry Fields framework, we have built an implementation of Blackbird using Python 3 as the embedding language --- an `embedded' DSL. This `Python-enhanced' Blackbird language provides the same core operations and follows the same grammar and syntactical rules as the standalone DSL, but, by nature, may also contain valid Python constructs. Furthermore, Strawberry Fields' `Python-enhanced' Blackbird provides users with additional quantum operations that are decomposed into lower-level Blackbird assembly commands.
We will introduce the elements of Blackbird through a series of basic examples, discussing more technical aspects as they arise.

\subsection{Operations}

Quantum computations consist of four main ingredients: state preparations, application of gates, performing measurements, and adding/removing subsystems. In Blackbird, these are all considered as \emph{Operations}, and share the same basic syntax. In the following code examples, we use the variable \verb+q+ for a set of qumodes (more specifically, a \emph{quantum register}), the details of which are deferred until the next section.

Our first considered Operation is state preparation. By default, qumodes are initialized in the vacuum state. Various other important CV states can be created with simple Blackbird commands. 

\begin{code}
	\pythonfile{code/states.py}
	\caption{Blackbird code for creating various CV quantum states.}
\end{code}

\noindent Blackbird state preparations such as those used in Codeblock 1 implicitly reset the existing state of the qumodes. Conceptually, the vertical bar symbol `$|$' separates Operations -- like state preparation -- from the registers that they act upon. Notice that we can use Operations inline, or construct them separately and reuse them several times. 

After creating states, we will want to transform these using quantum gates.

\begin{code}
	\pythonfile{code/gates.py}
	\caption{Blackbird code for applying various CV gates.}
\end{code}

\noindent Blackbird supports all of the gates listed in the previous section as well as a number of composite gates, each of which can be decomposed using the universal gates. The supported composite gates are: controlled X (\verb+CXgate+), controlled Z (\verb+CZgate+), quadratic phase (\verb+Pgate+), and two-mode squeezing (\verb+S2gate+). A full list of gates currently supported in Blackbird can be found in Appendix \ref{app:gates_etc}.

Finally, we can specify measurement Operations using Blackbird.

\begin{code}
	\pythonfile{code/measurements.py}
	\caption{Blackbird code for carrying out CV measurements.}
\end{code}

\noindent Measurements have several effects. For one, the numerical result of the measurement is placed in a classical register. As well, the state of all remaining qumodes is projected to the (normalized) conditional state for that measurement value. Finally, the state of the measured qumode is reset to the vacuum state. This is the typical behaviour of photonic hardware, where measurements absorb all the energy of the measured qumode. 

\subsection{Running Blackbird Programs in Python}
Within Python, Blackbird programs are managed by an \emph{Engine}. The function \verb+Engine+ in the Strawberry Fields API will instantiate an Engine, returning both the Engine and its corresponding quantum register. The Engine is used as a Python context manager, providing a convenient way to encapsulate Blackbird programs.

\begin{code}
	\pythonfile{code/engine.py}
	\caption{Code for declaring and running Blackbird programs using the Strawberry Fields library.}
\end{code}

\noindent The above code example is runnable and carries out a complete quantum computation, namely the preparation and measurement of an EPR entangled state. Note that Operations can be declared outside of the Engine, but their action on the quantum registers must come within the Engine context. Also notice that our register has a length of 2, so any single-mode Operations must act on specific elements, i.e., \verb+q[i]+, while two-mode Operations can act on \verb+q+ directly. Finally, the user must specify a backend -- as well as any backend-dependent settings -- when calling \verb+eng.run()+. We will discuss the Strawberry Fields backends further in a later section.

\subsection{Quantum and Classical Registers}

When a Strawberry Fields Engine is constructed, the user must specify the number of qumode subsystems to begin with. This number is required for the initialization of the Engine, but may change within a computation (e.g., when temporary ancilla modes are used). Qumodes can be added/deleted by using the \verb+New+ and \verb+Del+ Operations. 

\begin{code}
	\pythonfile[firstline=8,firstnumber=1]{code/add_del.py}
	\caption{Adding and deleting qumode subsystems.}
	\label{cb:add_del}
\end{code}

\noindent An Engine maintains a unique numeric indexing for the quantum registers based on the order they were added. When a subsystem is deleted from the circuit, no further gates can act on that register.

As stated earlier, measurement Operations produce classical information (the measurement result) and manipulate the corresponding register. Compared to previously released quantum programming frameworks, such as ProjectQ, PyQuil, and Qiskit, Strawberry Fields has been designed so that the notation  \texttt{q[i]}, while principally denoting the quantum register, may also encapsulate classical information or a classical register. This behaviour is contextual, and unique to Strawberry Fields. For example, prior to measurement, \texttt{q[i]} simply references a quantum register. Once a measurement operation is performed, \texttt{q[i]} continues to represent a quantum register --- now reset to the vacuum state --- as well as storing the numerical value of the measurement, accessible via the attribute \texttt{q[i].val}. Note that this numerical value is only available if a computation has been run up to the point of measurement. 
We may also use a classical measurement result \emph{symbolically} as a parameter in later gates without first running the computation. To do this, we simply pass the measured register (e.g., \verb+q[i]+) explicitly as an argument to the required gate. As before, the Strawberry Fields quantum register object is contextual --- when passed as a gate argument, Strawberry Fields implicitly accesses the encapsulated classical register.

\begin{code}
	\pythonfile[firstline=8,lastline=20,firstnumber=1]{code/classical_reg.py}
	\caption{Evaluating measurement results numerically and using them symbolically.}
\end{code}

\noindent In quantum algorithms, it is common to process a measurement result classically and use the post-processed value as a parameter for further operations in a circuit. Strawberry Fields provides the \verb+convert+ decorator to transform a user-specified numerical function into one which acts on registers.

\begin{code}
	\pythonfile[firstline=23,firstnumber=1]{code/classical_reg.py}
	\caption{Symbolically processing a measured value before using it.}
\end{code}

\subsection{Post-selection}
The measurement Operations in Strawberry Fields are stochastic in nature, with outcomes determined by some underlying quantum probability distribution. Often it is convenient to select specific values for these measurements rather than sampling them. For instance, we might want to explore the conditional state created by a specific value, determine the measurement-dependent corrections we need to make in a teleportation circuit, or even design an algorithm which inherently contains post-selection. This functionality is supported in Strawberry Fields through the optional keyword argument \verb+select+, which can be supplied for any measurement Operation. The measurement outcome will then return exactly this value, while the remaining modes will be projected into the conditional state corresponding to this value\footnote{Users should be careful to avoid post-selection on measurement values which have no probability of occuring given the current circuit state. In this case, the expected behaviour of a backend is not defined.}.

\begin{code}
	\pythonfile{code/select.py}
	\caption{Selecting a specific desired measurement outcome.}
\end{code}

\subsection{Decompositions}
In addition to the core CV operations discussed above, Strawberry Fields also provides support for some important decompositions frequently used in quantum optics. These include the (a) Williamson decomposition \cite{williamson1936on}, for decomposing arbitrary Gaussian states to a symplectic transformation acting on a thermals state, (b) the Bloch-Messiah decomposition \cite{bloch1975canonical,braunstein2005squeezing,simon1994quantum}, for decomposing the action of symplectic transformations to interferometers and single-mode squeezing, and (c) the Clements decomposition \cite{clements2016optimal}, for decomposing multi-mode linear interferometers into arrays of beamsplitters and rotations of fixed depth. In all cases, the resulting decomposition into the universal CV gate set may be viewed via the engine method \texttt{eng.print\_applied()}. Strawberry Fields thus provides a natural environment for embedding graphs and matrices in quantum optical circuits, and viewing the resulting physical components.

\begin{code}
	\pythonfile{code/decompositions.py}
	\caption{Using the in-built Clements decomposition to decompose a $2\times 2$ \texttt{Interferometer} into beamsplitters and phase rotations.}
\end{code}

\subsection{Optimization and Quantum Machine Learning}
Strawberry Fields can perform quantum circuit simulations using both numerical and symbolic representations. Numerical computation is the default operating mode and is supported by all three supplied backends. Symbolic computation is enabled only for the TensorFlow backend. In this section, we outline the main TensorFlow functionalities accessible through the Strawberry Fields frontend interface. More details about the corresponding TensorFlow backend are discussed in the next section.
TensorFlow \cite{abadi2016tensorflow} models calculations abstractly using a \emph{computational graph}, where individual operations are represented as nodes and their dependencies by directed edges. This viewpoint separates the symbolic representation of a computation from its numerical evaluation, and makes optimization and machine learning more amenable. On top of this, TensorFlow provides a number of advanced functionalities, including automatic gradient computation, GPU utilization, built-in optimization algorithms, and various other machine learning tools. Note that the term `quantum machine learning' will be used here in a hybrid sense, i.e., applying conventional machine learning methods to quantum systems.

To build a TensorFlow computational graph using Strawberry Fields, we instantiate an Engine, declare a circuit in Blackbird code, then execute \verb+eng.run+ on the TensorFlow (\verb+"tf"+) backend. To keep the underlying simulation fully symbolic, the extra argument \verb+eval=False+ must be given.

\begin{code}
	\pythonfile[lastline=16]{code/symbolic.py}
	\caption{Creating a TensorFlow computational graph for a quantum circuit.}
\end{code}

\noindent When we do this, any registers measured in the circuit will be populated with unevaluated \verb+Tensor+ objects rather than numerical values (without \verb+eval=False+, the TensorFlow backend returns purely numerical results, similar to the other simulators). 
These Tensors can still be evaluated numerically by running them in a TensorFlow \verb+Session+. In this case, measurement results will be determined stochastically on each evaluation.

\begin{code}
	\pythonfile[firstline=18,lastline=24]{code/symbolic.py}
	\caption{Numerically evaluating Tensors.}
\end{code}

When specifying a circuit in Blackbird, we can make use of various special symbolic TensorFlow classes, such as \verb+Variable+, \verb+Tensor+, \verb+placeholder+, or \verb+constant+.

\begin{code}
	\pythonfile[firstline=26,lastline=46]{code/symbolic.py}
	\caption{Using abstract TensorFlow classes as circuit parameters.}
\end{code}

\noindent In the above example, we supplied an additional \verb+feed_dict+ argument when evaluating. This is a Python dictionary which specifies the numerical values (typically, coming from a dataset) for every \verb+placeholder+ that appears in a circuit. However, as can be seen from the example, it is also possible to substitute desired values for other nodes in the computation, including the values stored in quantum registers. This allows us to easily post-select measurement values and explore the resulting conditional states. 

We can also perform post-processing of measurement results when working with TensorFlow objects. In this case, the functions decorated by \verb+convert+ should be written using TensorFlow \verb+Tensors+, \verb+Variables+, and operations.

\begin{code}
	\pythonfile[firstline=48,lastline=66]{code/symbolic.py}
	\caption{Processing a measurement result using a neural network. For compactness, the example uses a width 1 perceptron, but any continuous processing function supported by TensorFlow can be used.}
\end{code}
The TensorFlow backend additionally supports the use of batched processing, allowing for many evaluations of a quantum circuit to potentially be computed in parallel. Scalars are automatically broadcast to the specified batch size.
Finally, we can easily run circuit simulations on special-purpose hardware like GPUs or TPUs. 

\begin{code}
	\pythonfile[firstline=68]{code/symbolic.py}
	\caption{Running a batched computation and explicitly placing the computation on a GPU.}
\end{code}
\noindent By taking advantage of these additional functionalities of the TensorFlow backend, we can straightforwardly perform optimization and machine learning on quantum circuits in Strawberry Fields \cite{arrazola2018machine,quesada2018gaussianA}. A complete code example for optimization of a quantum circuit is located in Appendix \ref{app:algos}.

\section{Strawberry Fields' Quantum Simulators}
The ultimate goal is for Blackbird programs to be carried out on photonic quantum computers. To lay the groundwork for these emerging devices, Strawberry Fields comes with a suite of three CV quantum computer simulators specially designed for the CV model.
These simulators target different use-cases and support different functionality. 
For example, many important algorithms in the CV formalism involve only Gaussian states, operations, and measurements. We can take advantage of this structure to more efficiently simulate such computations. Other circuits have inherently non-Gaussian elements to them; for these, the Fock basis provides the standard description. These representations are available in Strawberry Fields in the \emph{Gaussian backend} and the \emph{Fock backend}, respectively. The third built-in backend is the \emph{TensorFlow backend}. Also using the Fock representation, this backend is geared primarily towards optimization and machine learning applications. 

Most Blackbird operations are supported across all three backends. A small subset, however, are not supported uniformly due to mathematical incompatibility. For example, the Gaussian backend does not generally support non-Gaussian gates or the preparation/measurement of Fock states. Sometimes it is also useful to work with a backend directly. To allow this, Strawberry Fields provides a backend API, giving access to additional methods and properties which are not part of the streamlined frontend API. A standalone backend can be created in Strawberry Fields using \verb+strawberryfields.backend.load_backend(name)+ where \verb+name+ is one of \verb+"gaussian"+, \verb+"fock"+, or \verb+"tf"+ (for comparison, the backend associated to an Engine \verb+eng+ is available via \verb+eng.backend+). 

Three important backend methods, common to all simulators, are \verb+begin_circuit+, \verb+reset+, and \verb+state+. The first command instantiates the circuit simulation, the second resets the simulation back to an initial vacuum state, clearing all previous operations, and the third returns a class which encapsulates the current quantum state of the simulator. In addition to containing the numerical (or symbolic) state data, state classes also contain a number of useful methods and attributes for further exploring the quantum state, such as \verb+fidelity+, \verb+mean_photon+, or \verb+wigner+. As a convenience for the user, all simulations carried out via \verb+eng.run+ will return a state class representing the final circuit state (see Appendix \ref{app:algos} for examples).

\subsection{Gaussian Backend}
This backend, written in NumPy, uses the symplectic formalism to represent CV systems. At a high level, this representation tracks the quantum state of an $N$-mode quantum system using two Gaussian components: a $2N$-dimensional displacement vector $\bar{ \bm{r}}$ and a $2N\times2N$-dimensional covariance matrix $\bm{V}$ (a deeper technical overview is located in Appendix \ref{app:technical_details}). After we have created a Gaussian state (either via \verb+state = eng.run(backend="gaussian")+ or by directly calling the \verb+state+ method of a Gaussian backend), we can access $\bar{ \bm{r}}$ and $\bm{V}$ via \verb+state.means+ and \verb+state.cov+, respectively. Other useful Gaussian state methods are \verb+displacement+ and \verb+squeezing+, which return the Gaussian parameters associated to the underlying state. 

The scaling of the symplectic representation with the number of modes is $\mathcal{O}(N^2)$. On one hand, this is quite powerful. It allows us to to efficiently simulate any computations which are fully Gaussian. On the other, the formalism is not expressive enough to simulate more general quantum computations. Only a small number of non-Gaussian methods are available for this backend. These are auxiliary methods where we extract some non-Gaussian information from a Gaussian state, but do not update the state of the circuit. One such method is \verb+fock_prob+, which is implemented using an optimized -- yet still exponentially scaling -- algorithm. This method enables simulation of the \emph{Gaussian boson sampling} algorithm \cite{hamilton2017gaussian} using the Gaussian backend; see Appendix \ref{app:algos} for a complete code example. 

\subsection{Fock Backend}
This backend, also written in NumPy, uses a fundamentally different description for qumodes than the Gaussian representation. As discussed in the introductory sections, the Fock representation encodes quantum computation in a countably infinite-dimensional Hilbert space. This representation is faithful, allowing a precise description of CV systems, in particular non-Gaussian circuits. 
Yet simulating infinite-dimensional systems leads to some computational tradeoffs which are not present for qubit simulators. Most importantly, we impose a \emph{cutoff dimension $D$} for simulations (chosen by the user), so the Fock backend only covers a restricted set of number states $\{\ket{0},...,\ket{D-1}\}$ for each mode. The size of simulated quantum systems thus depends on both the number of subsystems $N$ and the cutoff, being $\mathcal{O}(D^N)$. We contrast this with qubit systems, where the base is fixed, i.e., $\mathcal{O}(2^N)$. While these scalings are both exponential, in practice simulating qumode systems for $D>2$ is more computationally demanding than qubits. This is because the (truncated) qumode subsystems have a higher dimension and thus encode more information than their qubit counterparts.

Physically, imposing a cutoff is a reasonable strategy since higher photon-number states must have higher energy and, in practice, quantum-optical systems will have bounded energy (e.g., limited by the power of a laser). On the other hand, there are certainly states which can be easily prepared in the lab, yet would not fit accurately on the simulator. Thus, some care must be taken to trade off between the numerical cutoff value, the number of modes, and the energy scale of the circuit. If the energy scale is sufficiently low that all states fit within the specified cutoff, then simulations with the Fock and Gaussian backends will be in numerical agreement. 

Like the Gaussian backend, the Fock backend has a \verb+state+ method which encapsulates the numerical state, while also providing a number of methods and attributes specific to the Fock representation (such as \verb+ket+, \verb+trace+, and \verb+all_fock_probs+.). Unlike the Gaussian representation, mixed state simulations take up more resources than pure states. Pure states are represented in the Fock backend by an $D^N$-dimensional complex vector and mixed states by a $D^{2N}$-dimensional density matrix. Because of this extra overhead, by default the Fock backend will internally represent a quantum circuit as long as possible as a pure state, switching to the mixed state representation only when it becomes necessary. Most importantly, for $N>2$ qumodes, \emph{all state preparation Operations} (\verb+Vacuum+, \verb+Squeezed+, \verb+Fock+, etc.) \emph{cause the representation to become mixed}. This is because the mode where the state is prepared could be entangled with other modes. To keep physically consistent, the Fock backend will first trace out the relevant mode, necessitating a mixed state representation. When possible, \emph{it is recommended to apply gates to the (default) vacuum state in order to efficiently prepare pure states}. If desired, a mixed state simulation can be enforced by passing the argument \verb+pure=False+ when calling \verb+begin_circuit+.

\subsection{TensorFlow Backend}

The other built-in backend for Strawberry Fields is coded using TensorFlow. As a simulator, it uses the same internal representation as the Fock backend (Fock basis, finite cutoff, pure vs. mixed state representations, etc.) and has the same methods. It can operate as a numerical simulator similar to the other backends. Its main purpose, however, is to leverage the many powerful tools provided by TensorFlow to enable optimization and machine learning on quantum circuits. Much of this functionality was presented in the previous section, so we will not repeat it here. 

Like the other simulators, users can query the TensorFlow backend's \verb+state+ method to access the internal representation of a circuit's quantum state. This functions similarly to the Fock backend's \verb+state+ method, except that the state returned can be an unevaluated Tensor object when the keyword argument \verb+eval+ is set to \verb+False+. This state Tensor can be combined with any supported TensorFlow operations (\verb+norm+, \verb+self_adjoint_eig+, \verb+inv+, etc.) to enable optimization and machine learning on various properties of quantum circuits and quantum states. 

\section{Conclusions}
We have introduced Strawberry Fields, a multi-faceted software platform for continuous-variable quantum computing. The main components of this library -- a custom quantum programming language (Blackbird), a compiler engine, and a suite of quantum simulators targeting distinct applications -- have been presented in detail. Further information is available in both the Appendices and the Strawberry Fields online documentation. 
\newpage
\noindent The stage is now set for the broader community to use Strawberry Fields for exploration, research, and development of new quantum algorithms, specialized circuits, and machine learning models. We anticipate the creation of further software applications and backend modules for the Strawberry Fields platform (developed both internally and externally), providing advanced functionality and applications for quantum computing and quantum machine learning. 

\section{Acknowledgements}
We thank our colleagues at Xanadu for testing Strawberry Fields, reviewing this white paper, and providing helpful feedback. In particular, we thank Patrick Rebentrost for valuable discussions and suggestions.
\appendix

\clearpage
\section{Appendix A: The CV model}
\label{app:technical_details}
In this Appendix, we provide more technical and mathematical details about the CV model, the quantum computing paradigm underlying Strawberry Fields.

\subsection{Universal Gates for CV Quantum Computing}
In discrete qubit systems, the notion of a \emph{universal gate set} has the following meaning: given a set of universal gates, we can approximate an arbitrary unitary by composition of said gates.
In the CV setting, we have a similar situation: we can approximate a broad set of \emph{generators} -- i.e., the Hamiltonians appearing in Eq. (\ref{eq:unitary_generator}) -- by combining elements of a CV universal gate set. 
However, unlike the qubit case, we do not try to approximate all conceivable unitaries. Rather, we seek to create all generators that are a \emph{polynomial} function of the quadrature (or mode) operators of the system \cite{braunstein2005quantum,lloyd1999quantum}. Remember that generators of second-degree or lower belong to the class of \emph{Gaussian} operations, while all higher degrees are \emph{non-Gaussian}.

We can create a higher-order generator out of lower-order generators $\hat{A}$ and $\hat{B}$ by using the following two concatenation identities \cite{lloyd1999quantum}:
\begin{subequations}
\label{eq:comp}
\begin{eqnarray}
e^{-i \hat A \delta t} e^{-i \hat B \delta t} e^{i \hat A \delta t} e^{i \hat B \delta t} &=& e^{[\hat A, \hat B]\delta t^2}+O(\delta t^3), \label{comm}\\
e^{i \hat A \delta t/2} e^{i \hat B \delta t/2}  e^{i \hat B \delta t/2} e^{i \hat A \delta t/2} &=&e^{i (\hat A+\hat B)\delta t}+O( \delta t^2).
\end{eqnarray}
\end{subequations}
If we have two second-degree generators, such as $\hat u=\x^2+\p^2$ (the generator for the rotation gate) and $\hat s = \x \p+\p \x$ (the generator for the squeezing gate), and a third-degree (or higher) generator, such as $\x^3$ (the generator for the cubic phase gate), we can easily construct generators of all higher-degree polynomials, e.g., $\x^4 = - [\x^3,[\x^3,\hat u]]/(18 \hbar^2)$.
This reasoning can be extended by induction to any finite-degree polynomial in $\x$ and $\p$ (equivalently, in $\a$ and $\ad$) \cite{braunstein2005quantum,lloyd1999quantum}. 

In the above argument, it is important that at least one of the generators is third-degree or higher. Indeed, commutators of second-degree polynomials of $\x$ and $\p$ are also second-degree polynomials and thus their composition using Eq. (\ref{eq:comp}) cannot generate higher-order generators. The claim can be easily extended to $N$-mode systems and multivariate polynomials of the operators
\begin{align}
\label{eq:vecx}
\hat{\bm{ r}}=(\x_1, \ldots, \x_N, \p_1, \ldots, \p_N). 
\end{align}
Combining single-mode universal gates (including at least one of third-degree or higher) with some multimode interaction, e.g., the beamsplitter interaction generated by $\hat b_{i,j}=\p_i \x_j -\x_i \p_j$, we can construct arbitrary-degree polynomials of the quadrature operators of an $N$-mode system.

With the above discussion in mind, we can combine the set of single-qumode gates generated by $\{\x_i, \x_i^2, \x_i^3, \hat u_i \}$ and the two-mode gate generated by $\hat b_{i,j}$ for all pairs of modes into a universal gate set. The first, third and fourth single-mode generators correspond to the displacement, cubic phase and rotation gates in Table \ref{box:gates}, while the two-mode generator corresponds to the beamsplitter gate. Finally, the $\x^2$ generator corresponds to a quadratic phase gate, $\hat P(s) = \exp(i s \x^2/(2 \hbar))$. This gate can be written in terms of the single-mode squeezing gate and the rotation gate as follows: $\hat P(s) = \hat R(\theta) \hat S( r e^{i \phi})$, where $\cosh(r) = \sqrt{1+(s/2)^2}, \ \tan(\theta) = s/2, \phi = -\theta -\text{sign}(s) \pi/2$. Equivalently, we could have included the squeezing generator $\x \p + \p \x$ in place of the quadratic phase and still had a universal set. 

We have just outlined an efficient method to construct any gate of the form $\exp{(-iHt)}$, where the generator $H$ is a polynomial of the quadrature (or mode) operators. How can this be used for quantum computations? As shown in Eq. (\ref{eq:dichotomy}), the eigenstates of the $\x$ quadrature form an (orthogonal) basis for representing qumode states. Thus, these states are often taken as the computational basis of the CV model (though other choices are also available). By applying gates as constructed above and performing measurements in this computational basis (i.e., homodyne measurements), we can carry out a CV quantum computation. One primitive example \cite{braunstein2005quantum} is to compute the product of two numbers -- whose values are stored in two qumode registers -- and store the result in the state of a third qumode. Consider the generator $\x_1 \x_2 \p_3/\hbar$, which will cause the ``position'' operators to evolve according to
\begin{align}
\x_1 \to \x_1, \quad \x_2 \to \x_2, \quad \x_3 \to \x_3+\x_1 \x_2 t.
\end{align}
A measurement in the computational basis of mode 3 will reveal the value of the product $x_1 x_2$. Note that these encodings of classical continuous degress of freedom into quantum registers allows for the generalization of neural networks into the quantum regime \cite{killoran2018continuous}. In the following sections we show how more complicated algorithms are constructed.

\subsection{Multiport Gate Decompositions}
One important set of gates for which it is critical to derive a decomposition in terms of universal gates is the set of multiport interferometers. A multiport interferometer, represented by a unitary operator $\mathcal{\hat U}$ acting on $N$ modes, will map (in the Heisenberg picture) the annihilation operator of mode $i$ into a linear combination of all other modes
\begin{align}
\a_i \to \mathcal{\hat U}^\dagger \a_i \mathcal{\hat U} = \sum_{j} ( U_{i j} \a_j).
\end{align}
In order to preserve the commutation relations of different modes, the matrix $U$ must also be unitary $U U^\dagger =U^\dagger U=\I_N$. Note that every annihilation operator is mapped to a linear combination of annihilation operators and thus annihilation and creation operators are not mixed. Because of this, multiport interferometers generate all passive (in the sense that no photons are created or destroyed) linear optics transformations.
In a pioneering work by Reck \emph{et al.} \cite{reck1994experimental}, it was shown that any multiport interferometer can be realized using $N(N-1)/2$ beamsplitter gates distributed over $2N-3$ layers. 
Recently this result was improved upon by Clements \emph{et al.} \cite{clements2016optimal}, who showed that an equivalent decomposition can be achieved with the same number of beamsplitters but using only $N+1$ layers. 


\subsection{Gaussian Operations}
As mentioned in the previous subsections, generators that are at most quadratic remain closed under composition of their associated unitaries.  In the Heisenberg picture these quadratic generators will produce all possible linear transformations between the quadrature (or mode) operators,
\begin{align}
\a_i \to \sum_{j}  U_{i j} \a_j + V_{ij}\ad_j.
\end{align}
These operations are known as Gaussian operations. All gates in Table \ref{box:gates} are Gaussian operations except for the cubic phase gate. Pure Gaussian states are the set of states that can be obtained from the (multimode) vacuum state by Gaussian operations \cite{weedbrook2012gaussian,serafini2017quantum}. Mixed Gaussian states are obtained by applying Gaussian operations to thermal states or marginalizing pure Gaussian states. Analogous to Gaussian states, we can also define Gaussian measurements as the set of measurements whose positive-operator valued measure (POVM) elements can be obtained from vacuum via Gaussian transformations. Homodyne and heterodyne measurements are prominent examples of Gaussian measurements, whereas photon counting and photodetection are prominent examples of \emph{non}-Gaussian measurements.

An important result for the CV formalism is that Gaussian quantum computation, i.e., computation that occurs with Gaussian states, operations and measurements, can be \emph{efficiently} simulated on a classical computer (this is the foundation for the Gaussian backend in Strawberry Fields). This result is the CV version \cite{bartlett2002} of the Gottesman-Knill theorem of discrete-variable quantum information \cite{braunstein2005quantum}. Hence we need non-Gaussian operations in order to achieve quantum supremacy in the CV model. Interestingly, even in the restricted case where all states and gates are Gaussian, with only the final measurements being non-Gaussian, there is strong evidence that such a circuit cannot be efficiently simulated classically \cite{lund2014boson,hamilton2017gaussian}. More discussion and example code for this situation (known as \emph{Gaussian boson sampling}) is provided in Appendix \ref{app:algos}. 

\subsection{Symplectic Formalism}

In this section we review the symplectic formalism which lies at the heart of the Gaussian backend of Strawberry Fields. The symplectic formalism is an elegant and compact description of Gaussian states in terms of covariance matrices and mean vectors \cite{weedbrook2012gaussian,serafini2017quantum}. To begin, the commutation relations of the $2N$ position and momentum operators of Eq. (\ref{eq:vecx}) can be easily summarized as $[\hat r_i, \hat r_j] = \hbar i\Omega_{ij}$ where $\Omega=\left(
\begin{smallmatrix}
 0 & \I_N \\
- \I_N & 0
\end{smallmatrix}
\right)$ is the \emph{symplectic matrix}. Using the symplectic matrix, we can define the \emph{Weyl operator} $D(\bm{\xi})$ (a multimode displacement operator) and the \emph{characteristic function} $\chi(\bm{\xi})$ of a quantum $N$ mode state $\rho$:
\begin{align}
\hat D(\bm{\xi}) = \exp\left(i \hat{ \bm{r}} \Omega \bm{\xi}\right), \quad \chi(\bm{\xi}) = \langle \hat D(\bm{\xi}) \rangle_{\rho}, 
\end{align}
where $\bm{\xi} \in \mathbb{R}^{2N}$. We can now consider the Fourier transform of the characteristic function to obtain the \emph{Wigner function} of the state $\hat \rho$
\begin{align}
W(\bm{r}) = \int_{\mathbb{R}^{2N}} \frac{d^{2N} \xi}{(2 \pi)^{2N}} \exp( -i \bm{r} \Omega \bm{\xi} )\chi(\bm{\xi}).
\end{align}
The $2N$ real arguments $\bm{r}$ of the Wigner function are the eigenvalues of the quadrature operators from $\hat{\bm{r}}$. 

The above recipe maps an $N$-mode quantum state living in a Hilbert space to the real symplectic space $\mathcal{K}:=(\mathbb{R}^{2N},\Omega)$, which is called \emph{phase space}. The Wigner function is an example of a \emph{quasiprobability distribution}. Like a probability distribution over this phase space, the Wigner function is normalized to one; however, unlike a probability distribution, it may take negative values. Gaussian states have the special property that their characteristic function (and hence their Wigner function) is a Gaussian function of the variables $\bm{r}$. In this case, the Wigner function takes the form
\begin{align}
W(\bm{r}) =\frac{\exp\left( -\frac{1}{2} (\bm{r} -\bar{ \bm{r}}) \bm{V}^{-1} (\bm{r}-\bar{\bm{r}} ) \right)}{(2 \pi)^N \sqrt{\text{det} \bm{V}}}
\end{align}
where $\bar{ \bm{r}} = \langle \hat{\bm{r}} \rangle_\rho = \text{Tr}\left(\hat{ \bm{r}}  \hat \rho \right) $ is the displacement or mean vector and $\bm{V}_{ij} = \frac{1}{2} \langle \Delta r_i \Delta r_j+ \Delta r_i \Delta r_j \rangle_{\rho}$ with $\Delta \hat{\bm{ r}}=\hat{ \bm{r}}-\bar{ \bm{r}}$.
Note that the only pure states that have non-negative Wigner functions are the pure Gaussian states \cite{ferraro2005gaussian}. 

Each type of Gaussian state has a specific form of covariance matrix $\bm{V}$ and mean vector $\bar{\bm{r}}$. For the single-mode vacuum state, we have $\bm{V}=\frac{\hbar}{2}\I_2$ and $\bar{\bm{ r}}=(0,0)^T$. A thermal state (Eq. (\ref{eq:thermalstate})) has the same (zero) displacement but a covariance matrix $\bm{V}=(2 \bar n+1)\frac{\hbar}{2}\I_2$, where $\bar n$ is the mean photon number. A coherent state (Eq. (\ref{eq:coherentstate})), obtained by displacing vacuum, has the same $\bm{V}$ as the vacuum state but a nonzero displacement vector $\bar{ \bm{r}}=2 \sqrt{\frac{\hbar}{2}}(\Re(\alpha),\Im(\alpha))$. 
Lastly, a squeezed state (Eq. (\ref{eq:squeezedstate})) has zero displacement and covariance matrix  $\bm{V} = \frac{\hbar}{2} \text{diag}(e^{-2r},e^{2r})$.
In the limit $r \to \infty$, the squeezed state's variance in the $\x$ quadrature becomes zero and the state becomes proportional to the $\x$-eigenstate $\ket{x}$ with eigenvalue 0. Consistent with the uncertainty principle, the squeezed state's variance in $\p$ blows up. We can also consider the case $r \to -\infty$, where we find a state proportional to the eigenstate $\ket{p}$ of the $\p$ quadrature with eigenvalue 0. In the laboratory and in numerical simulation we must approximate every quadrature eigenstate using a finitely squeezed state (being careful that the variance of the relevant quadrature is much smaller than any other uncertainty relevant to the system). Any other quadrature eigenstate can be obtained from the $x=0$ eigenstate by applying suitable displacement and rotation operators. 
Finally, note that Gaussian operations will transform the vector of means via affine transformations and the covariance matrix via congruence transformations; for a detailed discussion of these transformations, see Sec. 2 of \cite{weedbrook2012gaussian}.

Given a $2N \times 2N$ real symmetric matrix, how can we check that it is a valid covariance matrix? If it is valid, which operations (displacement, squeezing, multiport interferometers) should be performed to prepare the corresponding Gaussian state?
To answer the first question: a $2N \times 2N$ real symmetric matrix $\tilde{ \bm{V}}$ corresponds to a Gaussian quantum state if and only if $\tilde{ \bm{V}}+i \frac{\hbar}{2}\Omega \geq 0$ (the matrix inequality is understood in the sense that the eigenvalues of the quantity $\tilde{ \bm{V}}+i \frac{\hbar}{2} \Omega$ are nonnegative). 
The answer to the second question is provided by the \emph{Bloch-Messiah reduction} \cite{bloch1975canonical,braunstein2005squeezing,simon1994quantum}. This reduction shows that any $N$-mode Gaussian state (equivalently any covariance matrix and vector of means) can be constructed by starting with a product of $N$ thermal states $\bigotimes_i \rho_i(\bar n_i)$ (with potentially different mean photon numbers), then applying a multiport interferometer $\mathcal{\hat U}$, followed by single-mode squeezing operations $\bigotimes_i S_i(z_i)$, followed by another multiport $\mathcal{\hat V}$, followed by single-mode displacement operations $\bigotimes_i D_i(\alpha_i)$. Explicitly,
\begin{align}
\rho_\text{Gaussian} &= \mathcal{\hat W} \left( \bigotimes_i \rho_i(\bar n_i) \right) \mathcal{\hat W}^\dagger, \label{decomp}\\
\mathcal{\hat W} &= \left(\bigotimes_i D_i(\alpha_i) \right)\mathcal{\hat V} \left(\bigotimes_i S_i(z_i) \right) \mathcal{\hat U}.
\end{align}


Note that if the Gaussian state is pure (which happens if and only if $\text{det}(\bm{V})=\left(\hbar/2 \right)^{2N}$), the occupation number of the thermal states in the Bloch-Messiah decomposition are all zero and the first interferometer will turn the vacuum to vacuum again. Thus for pure Gaussian states we need only generate $N$ single-mode squeezed states and send them through a single multiport interferometer $\mathcal{\hat V}$ before displacing. For a recent discussion of this decomposition see Ref. \cite{cariolaro2016bloch,cariolaro2016reexamination}. More generally, the occupation numbers of the different thermal  states in Eq. (\ref{decomp}) $n_i = (\nu_i-1)/2$ can be obtained by calculating the symplectic eigenvalues $\nu_i$ of the covariance matrix $\bm{V}$ . The symplectic eigenvalues come in pairs and are just the standard eigenvalues of the matrix $|i (2/\hbar)\Omega \bm{V}|$ where the modulus is understood in the operator sense (see Sec. II.C.1. of Ref. \cite{weedbrook2012gaussian}).

\clearpage
\onecolumngrid
\section{Appendix B: Strawberry Fields Operations}
\label{app:gates_etc}
\noindent In this Appendix, we present a complete list of the CV states, gates, and measurements available in Strawberry Fields. 

%

\begin{table}[htp!]
		\centering
		\rowcolors{1}{gray!10}{white}
		\arrayrulecolor{xgreen} 
		\setlength{\tabcolsep}{0pt}
		\setlength\extrarowheight{5pt}
		
		\begin{tabular}{m{0.333\textwidth}m{0.25\textwidth}m{0.416\textwidth}}
			\rowcolor{lightgreen}
			\textbf{~~~Operation}                   & \textbf{Name} & \textbf{Definition}\\
			\hline
			~~~\texttt{Vacuum()}         & {Vacuum state} & The vacuum state $\ket{0}$, representing zero photons \\[5pt]
			~~~\texttt{Coherent(a)} &   {Coherent state} & A displaced vacuum state,  $\ket{\alpha}=D(\alpha)\ket{0}$, $\alpha\in\mathbb{C}$\\[5pt]
			~~~\texttt{Squeezed(r,phi)}      & {Squeezed state} & A squeezed vacuum state, $\ket{z}=S(z)\ket{0}$,\newline where $z=re^{i\phi}$,~~ $r,\phi\in\mathbb{R}$, $r\geq 0$, $\phi\in[0,2\pi)$ \\[5pt]
			~~~\texttt{DisplacedSqueezed(a,r,phi)}	& {Displaced squeezed state} & A squeezed then displaced vacuum state,\newline  $\ket{\alpha,z}=D(\alpha)S(z)\ket{0}$	\\[5pt]
			~~~\texttt{Thermal(n)}     & {Thermal state} & $\rho(\bar{n})= \sum_{n=0}^\infty[\nbar^n/(1+\nbar)^{n+1}]\ketbra{n}{n}$,\newline where $\bar{n}\in\mathbb{R}^+$ is the mean photon number\\[5pt]
			~~~\texttt{Fock(n)}*     & {Fock state} or {number state} & $\ket{n}$, where $n\in\mathbb{N}_0$ represents the photon number \\[5pt]
			~~~\texttt{Catstate(a,p)}*     & Cat state & $\frac{1}{\sqrt{\mathcal{N}}}(\ket{\alpha}+\exp({i\pi p})\ket{-\alpha})$, where $p=0,1$ gives an even/odd cat state and $\mathcal{N}$ is the normalization\\[5pt]
			~~~\texttt{Ket(x)}*         & Arbitrary Fock-basis ket & Prepare an arbitrary multi-mode pure state, represented by array \texttt{x}, in the Fock basis.\\[5pt]
			~~~\texttt{DensityMatrix(x)}*         & Arbitrary Fock basis state & Prepare an arbitrary multi-mode mixed state, represented by a density matrix array \texttt{x} in the Fock basis.
		\end{tabular}
	\captionsetup{name=Table,type=table}
	\caption{State preparations available in Strawberry Fields. Those indicated with an asterisk (*) are non-Gaussian.} 
	\label{A2:tab:states}
\end{table}

\begin{table}[h!]
	\centering
	\rowcolors{1}{gray!10}{white}
	\arrayrulecolor{xgreen} 
	\setlength{\tabcolsep}{0pt}
	\setlength\extrarowheight{5pt}
	\begin{tabular}{m{0.27\textwidth}m{0.27\textwidth}m{0.459\textwidth}}
		\rowcolor{lightgreen}
		\textbf{~~~Operation}                   & \textbf{Name} & \textbf{Definition}\\                    
		\hline
		~~~\texttt{Dgate(a)}         & {Displacement gate} & $\begin{aligned}D(\alpha)&=\exp({\alpha \ad -\alpha^* \a})\\ &= \exp\left({-i (\Re(\alpha) \hat{p} -\Im(\alpha) \hat{x})\sqrt{{2}/{\hbar}}}\right), ~~\alpha\in\mathbb{C} \end{aligned}$\\[5pt]
		~~~\texttt{Xgate(x)} & {Position displacement gate} & $X(x) = D(x/\sqrt{2 \hbar})=\exp\left({-ix\hat{p}/\hbar}\right)$, $x\in\mathbb{R}$\\[5pt]
		~~~\texttt{Zgate(p)} & {Momentum displacement gate} & $Z(p) = D(ip/\sqrt{2 \hbar})=\exp\left({ip\hat{x}/\hbar}\right)$, $p\in\mathbb{R}$\\[5pt]
		~~~\texttt{Sgate(r,phi)} &{Squeezing gate} & $S(z) = \exp\left({(z^* {\a^2 -z \ad}^2)/2}\right)$, where \newline $z=r\exp\left({i\phi}\right)$, ~~$r,\phi\in\mathbb{R}$, $r\geq 0$, $\phi\in[0,2\pi)$\\[5pt]
		~~~\texttt{Rgate(theta)} & {Rotation gate} & $R(\theta) = \exp\left({i\theta \ad \a}\right)$, where $\theta\in[0,2\pi)$\\[5pt]
		~~~\texttt{Fouriergate()} & {Fourier gate} & $F=R(\pi/2) = \exp\left({i(\pi/2)\ad \a}\right)$\\[5pt]
		~~~\texttt{Pgate(s)} & {Quadratic phase gate}  & $P(s) = \exp\left({is\hat{x}^2/(2 \hbar)}\right)$, where $s\in\mathbb{R}$\\[5pt]
		~~~\texttt{Vgate(g)}* & {Cubic phase gate}  & $V(\gamma) = \exp\left({i\gamma\hat{x}^3/(3 \hbar)}\right)$, where $\gamma\in\mathbb{R}$\\[5pt]
		~~~\texttt{Kgate(k)}* & {Kerr interaction gate}  & $K(\kappa) = \exp\left({i\kappa\left(\ad\a\right)^2}\right)$, where $\kappa\in\mathbb{R}$
	\end{tabular}
	\captionsetup{name=Table,type=table}
	\caption{Single mode gate operations available in Strawberry Fields. Those indicated with an asterisk (*) are non-Gaussian and can only be used with a backend that uses the Fock representation.}
	\label{A2:tab:gates}
\end{table}

\begin{table}[h!]
		\centering
		\rowcolors{1}{gray!10}{white}
		\arrayrulecolor{xgreen} 
		\setlength{\tabcolsep}{0pt}
		\setlength\extrarowheight{5pt}
		\begin{tabular}{m{0.333\textwidth}m{0.333\textwidth}m{0.333\textwidth}}
			\rowcolor{lightgreen}
			\textbf{~~~Operation}                   & \textbf{Name} & \textbf{Definition} \\                    
			\hline
			~~~\texttt{BSgate(theta,phi)}         & {Beamsplitter} &  $B(\theta,\phi)=\exp\left(\theta (e^{i \phi} \a_1 \ad_2 -e^{-i \phi}\ad_1 \a_2) \right)$,\newline where the transmissivity and reflectivity \newline amplitudes are $t=\cos\theta$, $r=e^{i\phi}\sin\theta$\\
			~~~\texttt{S2gate(r,p)} & {Two-mode squeezing gate}& $S_2(z) =\exp\left({z^*\a_1\a_2-z\ad_1\ad_2}\right)$,\newline where $z=r\exp\left({i\phi}\right)$\\
			~~~\texttt{CXgate(s)} & {Controlled-X} or {addition gate}& $CX(s) = \exp\left({-is\hat{x}_1 \hat{p}_2/\hbar}\right)$, $s\in\mathbb{R}$\\
			~~~\texttt{CZgate(s)} & {Controlled phase shift gate}& $CZ(s) = \exp\left({is\hat{x}_1\hat{x}_2/\hbar}\right)$, $s\in\mathbb{R}$\\
			~~~\texttt{CKgate(k)}* & {Controlled Kerr interaction gate}& $CK(\kappa) = \exp\left({i\kappa \ad_1\a_1\ad_2\a_2}\right)$, $\kappa\in\mathbb{R}$
		\end{tabular}
	\captionsetup{name=Table,type=table}
	\caption{Two-mode gate operations available in Strawberry Fields.}
	\label{A2:tab:gates2}
\end{table}

\begin{table}[h!]
	\centering
	\rowcolors{1}{gray!10}{white}
	\arrayrulecolor{xgreen} 
	\setlength{\tabcolsep}{0pt}
	\setlength\extrarowheight{5pt}
	\begin{tabular}{m{0.28\textwidth}m{0.3\textwidth}m{0.42\textwidth}}
		\rowcolor{lightgreen}
		\textbf{~~~Operation}                   & \textbf{Name} & \textbf{Definition} \\                    
		\hline
		~~~\texttt{Gaussian(cov,mu)}         & {Gaussian state preparation} &  Prepares an arbitrary $N$-mode Gaussian state defined by covariance matrix $V\in\mathbb{R}^{2N\times 2N}$ and means vector $\mu\in\mathbb{R}^{2N}$ using the Williamson decomposition\\
		~~~\texttt{GaussianTransform(S)}         & {Gaussian transformation} &  Applies an $N$-mode Gaussian transformation defined by symplectic matrix $S\in\mathbb{R}^{2N\times 2N}$ using the Bloch-Messiah decomposition\\
		~~~\texttt{Interferometer(U)}         & {Multi-mode linear interferometer} &  Applies an $N$-mode interferometer defined by unitary matrix $U\in\mathbb{C}^{N\times N}$ using the Clements decomposition\\
	\end{tabular}
	\captionsetup{name=Table,type=table}
	\caption{Multi-mode Gaussian decompositions available in Strawberry Fields.}
	\label{A2:tab:decompositions}
\end{table}

\begin{table}[h!]
	\centering
	\rowcolors{1}{gray!10}{white}
	\arrayrulecolor{xgreen} 
	\setlength{\tabcolsep}{0pt}
	\setlength\extrarowheight{5pt}
	\begin{tabular}{m{0.333\textwidth}m{0.253\textwidth}m{0.413\textwidth}}
		\rowcolor{lightgreen}
		\textbf{~~~Operation}                   & \textbf{Name} & \textbf{Definition}\\                                  
		\hline
		~~~\texttt{MeasureHomodyne(phi)} & {Homodyne measurement}& Projects the state onto $|x_\phi\rangle\langle x_\phi|$\newline where $\hat{x}_\phi=\cos\phi\hat{x}+\sin\phi\hat{p}$\\
		~~~\texttt{MeasureHeterodyne()} & {Heterodyne measurement}& Projects the state onto the coherent states, sampling from the joint Husimi distribution $\frac{1}{\pi}\left\langle\alpha\middle|\rho\middle|\alpha\right\rangle$\\
		~~~\texttt{MeasureFock()}*         & {Photon counting}& Projects the state onto $|n\rangle\langle n|$
	\end{tabular}
	\captionsetup{name=Table,type=table}
	\caption{Measurement operations available in Strawberry Fields. Those indicated with an asterisk (*) are non-Gaussian and can only be used with a backend that uses the Fock representation.}
	\label{A2:tab:meas}
\end{table}

\FloatBarrier

\twocolumngrid
\subsection{Gate Decompositions}
\noindent In addition, the Strawberry Fields frontend can be used to provide decompositions of certain compound gates. The following gate decompositions are currently supported.

\subsubsection{Quadratic phase gate}
\noindent The quadratic phase shift gate \texttt{Pgate(s)} is decomposed into a squeezing and a rotation, $$P(s) = R(\theta) S(r e^{i \phi}),$$ where $\cosh(r) = \sqrt{1+(s/2)^2},$ $\tan(\theta) = s/2,$ and $\phi = -\text{sign}(s)\frac{\pi}{2} -\theta$.

\subsubsection{Two-mode squeeze gate}
\noindent The two-mode squeeze gate \texttt{S2gate(z)} is decomposed into a combination of beamsplitters and single-mode squeezers $$S_2(z) = B^\dagger(\pi/4,0)  \left[ S(z) \otimes S(-z)\right] B(\pi/4,0) $$

\vfill\null
	
\subsubsection{Controlled addition gate}
\noindent The controlled addition or controlled-X gate \texttt{CXgate(s)} is decomposed into a combination of beamsplitters and single-mode squeezers $$CX(s) = B(\phi,0) \left[S(r) \otimes S(-r) \right] B(\pi/2+\phi,0),$$ where $\sin(2 \phi) = -1/{\cosh (r)}$, $\cos(2 \phi)=-\tanh(r)$, and $\sinh(r) = -s/2$.

\subsubsection{Controlled phase gate}
\noindent The controlled phase shift gate \texttt{CZgate(s)} is decomposed into a controlled addition gate, with two rotation gates acting on the target mode, $$CZ(s) = \left[\I\otimes R(\pi/2)\right] CX(s)\left[\I\otimes R(\pi/2)^\dagger\right]$$


\clearpage
\twocolumngrid
\section{Appendix C: Quantum Algorithms}
\label{app:algos}
In this Appendix, we present full example code for several important algorithms, subroutines, and use-cases for Strawberry Fields. These examples are presented in more detail in the online documentation located at \href{https://strawberryfields.readthedocs.io}{strawberryfields.readthedocs.io}.

\subsection{Quantum Teleportation}


Quantum teleportation --- sometimes referred to as state teleportation to avoid confusion with gate teleportation --- is the reliable transfer of an unknown quantum state across spatially separated qubits or qumodes, through the use of a classical transmission channel and quantum entanglement \cite{bennett1993teleport}. Considered a fundamental quantum information protocol, it has applications ranging from quantum communication to enabling distributed information processing in quantum computation \cite{furusawa2011teleport}.

\begin{figure}[h]
	\mbox{\import{diagrams/}{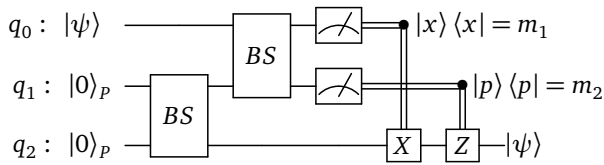}}
	\caption{State teleportation of state $\ket{\psi}$ from mode $q_0$ to mode $q_2$.}
        \label{fig:state_teleportation_circuit}
\end{figure}

In general, all quantum teleportation circuits work on the same basic principle. Two distant operators, Alice and Bob, share a maximally entangled quantum state (in discrete variables, any one of the four Bell states, and in CVs, a maximally entangled state for a fixed energy), and have access to a classical communication channel. Alice, in possession of an unknown state which she wishes to transfer to Bob, makes a joint measurement of the unknown state and her half of the entangled state, by projecting onto the Bell or quadrature basis. By transmitting the results of her measurement to Bob, Bob is then able to transform his half of the entangled state to an accurate replica of the original unknown state, by performing a conditional phase flip (for qubits) or displacement (for qumodes) \cite{steeb2006teleport}. The CV teleportation circuit is shown in Fig. \ref{fig:state_teleportation_circuit}.

\begin{code}
	\pythonfile{code/teleportation.py}
	\caption{Quantum teleportation of a coherent state in Strawberry Fields. Note: while the CV quantum teleportation algorithm relies on infinitely squeezed resource states, in practice a squeezing magnitude of -2 ($\sim18$ dB) is sufficient.}
\end{code}

\subsection{Gate Teleportation}

In the quantum state teleportation algorithm discussed in the previous section, the quantum state is transferred from the sender to the receiver. However, quantum teleportation can be used in a much more powerful manner, by simultaneously transforming the teleported state --- this is known as gate teleportation. 

In gate teleportation, rather than applying a quantum unitary directly to the state prior to teleportation, the unitary is applied indirectly, via the projective measurement of the first subsystem onto a particular basis. This measurement-based approach provides significant advantages over applying unitary gates directly, for example by reducing resources, and in the application of experimentally hard-to-implement gates \cite{furusawa2011teleport}. In fact, gate teleportation forms a universal quantum computing primitive, and is a precursor to cluster state models of quantum computation \cite{gu2009gate}. 

\begin{figure}
	\includegraphics[scale=1]{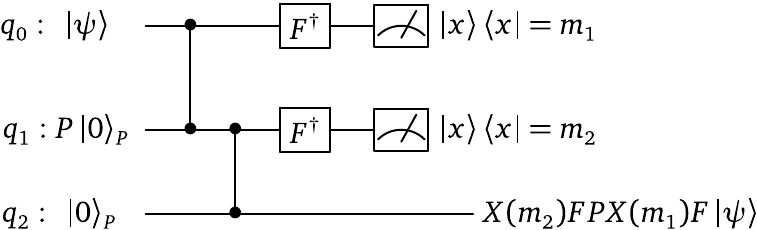}
	\caption{Gate teleportation of a quadratic phase gate $P$ onto input state $\ket{\psi}$.}
        \label{fig:gate_teleportation_circuit}
\end{figure}

First described by \citet{gottesman1999gate} in the case of qubits, gate teleportation was generalised for the CV case by \citet{bartlett2003gate} (see Fig. \ref{fig:gate_teleportation_circuit}). In an analogous process to the discrete-variable case, we begin with the algorithm for local state teleportation. Unlike the spatially-separated quantum state teleportation we considered in the previous section, local teleportation can transport the state using only two qumodes; the state we are teleporting is entangled directly with the squeezed vacuum state in the momentum space through the use of a controlled phase gate. To recover the teleported state exactly, we must perform Weyl-Heisenberg corrections to the second mode; here, that would be $F^\dagger X(m)^\dagger$, where $m$ is the correction based on the Homodyne measurement. However, for convenience and simplicity, it is common to simply write the circuit without the corrections applied explicitly.
\begin{code}
	\pythonfile{code/gate_teleportation.py}
	\caption{Gate teleportation of a quadratic phase gate in Strawberry Fields.}
\end{code}
Note that additional gates can be added to the gate teleportation scheme described above simply by introducing additional qumodes with the appropriate projective measurements, all `stacked vertically' (i.e., coupled to each consecutive qumode via a conditional phase gate). It is from this primitive that the model of cluster state quantum computation can be derived \cite{gu2009gate}.

\subsection{Boson Sampling}
Introduced by \citet{aaronson2011computational}, boson sampling presented a slight deviation from the general approach in quantum computation. Rather than presenting a theoretical model of universal quantum computation (i.e., a framework that enables quantum simulation of any arbitrary Hamiltonian \cite{nielsen2002quantum}), boson sampling-based devices are instead an example of an intermediate quantum computer, designed to experimentally implement a computation that is intractable classically \cite{tillmann2013boson,spagnolo2014boson,crespi2013boson,spring2012boson}.

\begin{figure}
	\mbox{\import{diagrams/}{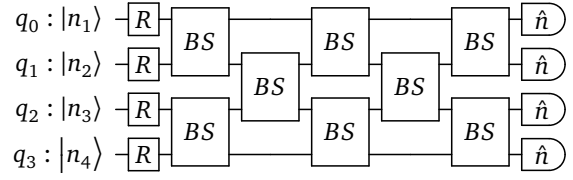}}
	\caption{4-mode boson sampling.}
	\label{fig:boson_sampling}
\end{figure}

Boson sampling proposes the following quantum linear optics scheme. An array of single photon sources is set up, designed to simultaneously emit single photon states into a multimode linear interferometer; the results are then generated by sampling from the probability of single photon measurements from the output of the linear interferometer.

For example, consider $N$ single photon Fock states, $\ket{\psi}=\ket{m_1,m_2,\dots,m_N}$, composed of $b=\sum_i m_i$ photons, incident on an $N$-mode linear interferometer described by the unitary $U$ acting on the input mode creation and annihilation operators. It was shown that the probability of detecting $n_j$ photons at the $j$th output mode is given by \cite{aaronson2011computational}
\begin{align}
\mathrm{Pr}(n_1,\dots, n_N) = \frac{\left|\text{Per}(U_{st})\right|^2}{m_1!\cdots m_N!n_1!\cdots n_N!};
\end{align}
i.e., the sampled single photon probability distribution is proportional to the permanent of $U_{st}$, a submatrix of the interferometer unitary, dependent upon the input and output Fock states. Whilst the determinant can be calculated efficiently on classical computers, calculation of the permanent belongs to the computational complexity class \#P-Hard problems \cite{valiant1979boson}, which are strongly believed to be classically hard to calculate. This implies that simulating boson sampling is an intractable task for classical computers, providing an avenue for the demonstration of quantum supremacy. 

Continuous-variable quantum computation is ideally suited to the simulation and demonstration of the boson sampling scheme, due to its grounding in quantum optics. In quantum linear optics, the multimode linear interferometer is commonly decomposed into two-mode beamsplitters (\texttt{BSgate}) and single-mode phase shifters (\texttt{Rgate}) \cite{reck1994experimental}, allowing for a straightforward translation into a CV quantum circuit. In order to allow for arbitrary linear unitaries on $m$ imput modes, we must have a minimum of $m+1$ columns in the beamsplitter array \cite{clements2016optimal}. 
\begin{code}
	\pythonfile{code/boson_sampling.py}
	\caption{$4$-mode boson sampling example in Strawberry Fields. Parameters are chosen arbitrarily.}
\end{code}

\subsection{Gaussian Boson Sampling}
\begin{figure}
	\mbox{\import{diagrams/}{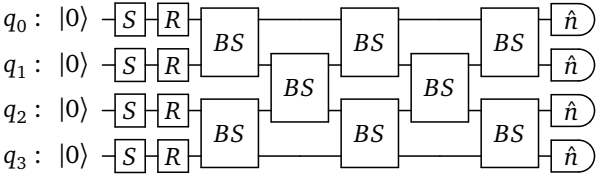}}
	\caption{4-mode Gaussian boson sampling.}
	\label{fig:gaussian_boson_sampling}
\end{figure}
While boson sampling allows the experimental implementation of a sampling problem that is hard to simulate classically, one of the setbacks in experimental setups is scalability, due to its dependence on an array of simultaneously emitting single photon sources. Currently, most physical implementations of boson sampling make use of a process known as Spontaneous Parametric Down-Conversion (SPDC) to generate the single-photon source inputs. However, this method is non-deterministic -- as the number of modes in the apparatus increases, the average time required until every photon source emits a simultaneous photon increases exponentially.

In order to simulate a deterministic single photon source array, several variations on boson sampling have been proposed; the most well-known being scattershot boson sampling \cite{lund2014boson}. However, a recent boson sampling variation by \cite{hamilton2017gaussian} negates the need for single photon Fock states altogether, by showing that incident Gaussian states -- in this case, single mode squeezed states -- can produce problems in the same computational complexity class as boson sampling. Even more significantly, this mitigates the scalability problem with single photon sources, as single mode squeezed states can be simultaneously generated experimentally.

With an input ensemble of $N$ single-mode squeezed states with squeezing parameter $z=r\in\mathbb{R}$, incident on a linear-interferometer described by unitary $U$, it can be shown that the probability of detecting an output photon pattern $(n_1, \dots, n_N)$, where $n_k\in\{0,1\}$, is given by \cite{hamilton2017gaussian}
\begin{align}
	\mathrm{Pr}(n_1,\dots,n_N) = \frac{\left|\text{Haf}[(U\bigoplus_i\tanh(r_i)U^T)_S]\right|^2}{ \prod_{i} \cosh(r_i)},
\end{align}
where $S$ denotes the subset of modes where a photon was detected and $\mathrm{Haf}[\cdot]$ is the \emph{Hafnian} \cite{caianiello1973combinatorics, barvinok2016combinatorics}. That is, the sampled single photon probability distribution is proportional to the hafnian of a submatrix of $U\bigoplus_i\tanh(r_i)U^T$. The hafnian is known to be a \textit{generalization} of the permanent, and can be used to count the number of perfect matchings of an arbitrary graph \cite{bradler2017gaussian}.
The formula above can be generalized to pure Gaussian states with finite displacements by using the loop hafnian function which counts the number of perfect matchings of graphs with loops\cite{quesada2018faster}.
Since any algorithm that could calculate the hafnian could also calculate the permanent, it follows that calculating the hafnian remains a classically hard problem; indeed, the best known classical algorithm for the calculation of a hafnian of an arbitrary symmetric complex matrix of size $N$ scales like $O(N^3 2^{N/2})$ \cite{bjorklund2018faster}. The hardness of approximate GBS, under imperfections such as loss, is a subject of current research \cite{quesada2018gaussian,gupt2018classical}.
\begin{code}
	\pythonfile{code/gaussian_boson_sampling.py}
	\caption{$4$-mode Gaussian boson sampling example in Strawberry Fields. Parameters are chosen arbitrarily.}
\end{code}

\subsection{Instantaneous Quantum Polynomial (IQP)}
Like boson sampling and Gaussian boson sampling before it, the instantaneous quantum polynomial (IQP) protocol is a restricted, non-universal model of quantum computation, designed to implement a computation that is classically intractable and verifiable by a remote adjudicator. First introduced by \citet{shepherd2009iqp}, IQP circuits are defined by the following conditions: (i) there are $N$ inputs in state $\ket{0}$ acted on by Hadamard gates, (ii) the resulting computation is diagonal in the computational basis by randomly selecting from the set $U=\{R(\pi/4),\sqrt{CZ}\}$ (hence the term `instantaneous'; since all gates commute, they can be applied in any temporal order), and (iii) the output qubits are measured in the Pauli-X basis. Efficient classical sampling of the resulting probability distribution $H^{\otimes N}UH^{\otimes N}\ket{0}^{\otimes N}$ --- even approximately \cite{bremner2016average} or in the presence of noise \cite{bremner2017achieving} --- has been shown to be \#P-hard, and would result in the collapse of the polynomial hierarchy to the third level \cite{lund2017iqp,bremner2010classical}.

Unlike boson sampling and Gaussian boson sampling, however, the IQP protocol was not constructed with quantum linear optics in mind. Nevertheless, the IQP model was recently extended to the CV formulation of quantum computation by \citet{douce2017iqp}, taking advantage of the ability to efficiently prepare large resource states, and the higher efficiencies afforded by homodyne detection. Furthermore, the computational complexity results of the discrete-variable case apply equally to the so-called CV-IQP model, assuming a specific input squeezing parameter dependent on the circuit size. The CV-IQP model is defined as follows:
\begin{enumerate}
	\item inputs are squeezed momentum states $\ket{z}$, where $z=r\in\mathbb{R}$ and $r<0$;
	\item the unitary transformation is diagonal in the $\hat{x}$ quadrature basis, by randomly selecting from the set of gates $U=\{Z(p),CZ(s),V(\gamma)\}$;
	\item and finally, homodyne measurements are performed on all modes in the $\hat{p}$ quadrature.
\end{enumerate}

\begin{figure}
	\mbox{\import{diagrams/}{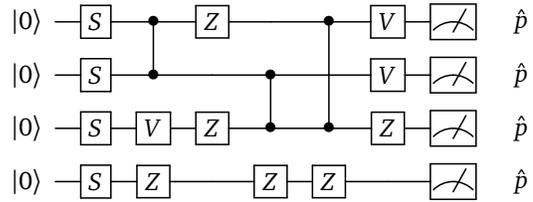}}
	\caption{4-mode instantaneous quantum polynomial (IQP) CV circuit example.}
\end{figure}

\begin{code}
	\pythonfile{code/IQP.py}
	\caption{4-mode instantaneous quantum polynomial (IQP) example in Strawberry Fields.}
\end{code}

Moreover, the resulting probability distributions have been shown to be given by integrals of oscillating functions in large dimensions, which are believed to be intractable to compute by classical computers. This leads to the result that even in the case of finite squeezing and reduced measurement precision, approximate sampling from the output CV-IQP model remains classically intractable \cite{douce2017iqp, arrazola2017quantum}.


\subsection{Hamiltonian Simulation}
The simulation of atoms, molecules and other biochemical systems is another application uniquely suited to quantum computation. For example, the ground state energy of large systems, the dynamical behaviour of an ensemble of molecules, or complex molecular behaviour such as protein folding, are often computationally hard or downright impossible to determine via classical computation or experimentation \cite{aspuruguzik2005hamiltonian,whitfield2011hamiltonian}. 

In the discrete-variable qubit model, efficient methods of Hamiltonian simulation have been discussed at length, providing several implementations depending on properties of the Hamiltonian, and resulting in a linear simulation time \cite{childs2012hamiltonian,berry2006hamiltonian}. Efficient implementations of Hamiltonian simulation also exist in the CV formulation \cite{kalajdzievski2018hamiltonian}, with specific application to Bose-Hubbard Hamiltonians (describing a system of interacting bosonic particles on a lattice of orthogonal position states \cite{sowinski2012hamiltonian}). As such, this method is ideally suited to photonic quantum computation.

\begin{figure}[h]
	\mbox{\import{diagrams/}{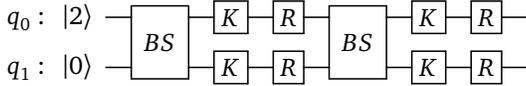}}
	\caption{Bose-Hubbard Hamiltonian simulation for a 2-node lattice, with $k=2$. The error is of the order $\mathcal{O}(t^2/k)$.}
\end{figure}

Considering a lattice composed of two adjacent nodes characterised by Adjacency matrix $A$, the Bose-Hubbard Hamiltonian is given by 
\begin{align}
	H&= J\sum_{i}\sum_j A_{ij} \ad_i\a_j + \frac{1}{2}U\sum_i \hat{n}_i(\hat{n}_i-1)\notag\\
	 &= J(\ad_1 \a_2 + \ad_2\a_1) + \frac{1}{2}U (  \hat{n}_1^2 - \hat{n}_1 + \hat{n}_2^2 - \hat{n}_2),
\end{align}
where $J$ represents the transition of the boson between nodes, and $U$ is the on-site interaction potential. Applying the Lie product formula, we find that
\begin{align}
	e^{iHt} &= \left[\exp\left({-i\frac{ J t}{k}(\ad_1 \a_2 + \ad_2\a_1)}\right)\exp\left(-i\frac{Ut}{2k}\hat{n}_1^2\right)\right.\notag\\
	&~~~\left.\exp\left(-i\frac{Ut}{2k}\hat{n}_2^2\right)\exp\left(i\frac{Ut}{2k}\hat{n}_1\right)\exp\left(i\frac{Ut}{2k}\hat{n}_2\right)\right]^k\notag\\
	&~~~+\mathcal{O}\left(t^2/k\right),
\end{align}
where $\mathcal{O}\left(t^2/k\right)$ is the order of the error term, derived from the Lie product formula. Comparing this to the form of various gates in the CV circuit model, we can write this as the product of beamsplitters, Kerr gates, and rotation gates:
\begin{align}
	e^{iHt} &= \left\{BS\left(\theta,\phi\right)\left[K(r)R(-r)\otimes K(r)R(-r)\right]\right\}^k+\mathcal{O}\left(t^2/k\right)
\end{align}
where $\theta=-Jt/k$, $\phi=\pi/2$, and $r=-Ut/2k$. Using $J=1$, $U=1.5$, $k=20$, and a timestep of $t=1.086$, this can be easily implemented in Strawberry Fields using only $20\times 3=60$ gates.
\begin{code}
	\pythonfile{code/hamiltonian_simulation.py}
	\caption{Tight-binding Hamiltonian simulation for a 2-node lattice in Strawberry Fields.}
\end{code}
For more complex Hamiltonian CV decompositions, including those with nearest-neighbour, see
\citet{kalajdzievski2018hamiltonian}. This decomposition is also implemented in the SFOpenBoson plugin \cite{mcclean2017openfermion}, which provides an interface between OpenFermion, the quantum electronic structure package, and Strawberry Fields. This allows arbitrary Bose-Hubbard Hamiltonians, generated in OpenFermion, to be simulated using Strawberry Fields.

\subsection{Optimization of Quantum Circuits}

One of the unique features of Strawberry Fields is that it has been designed from the start to support modern computational methods like automatic gradients, optimization, and machine learning by leveraging a TensorFlow backend. We have already given an overview in the main text of how these features are accessed in Strawberry Fields. We present here a complete example for the optimization of a quantum circuit. Our goal in this circuit is to find the Dgate parameter which leads to the highest probability of a $n=1$ Fock state output. This simple baseline example can be straightforwardly extended to much more complex circuits. As optimization is a key ingredient of machine learning, this example can also serve as a springboard for more advanced data-driven modelling tasks. 

We note that optimization here is in a \emph{variational} sense, i.e., we choose an ansatz for the optimization by fixing the discrete structure of our circuits (the selection and arrangement of specific states, gates, and measurements). The optimization then proceeds over the parameters of these operations.  Finally, we emphasize that for certain circuits and objective functions, the optimization might be non-convex in general. Thus we should not assume (without proof) that the solution obtained via a Strawberry Fields optimization is always the global optimum. However, it is often the case in machine learning that local optima can still provide effective solutions, so this may not be an issue, depending on the application.

\vfill\null

\begin{code}
	\pythonfile{code/optimization.py}
	\caption{Example of optimizing quantum circuit parameters in Strawberry Fields. This can be extended to a machine learning setting by incorporating data and a more sophisticated loss function.}
\end{code}

\bibliographystyle{unsrtnat}
\bibliography{references}

\end{document}